\documentclass[12pt]{article}
\usepackage{amssymb,amscd,array}
\catcode `\@=11
\@addtoreset{equation}{section}

\def\harr#1#2{\smash{\mathop{\hbox to .5in{\rightarrowfill}}
\limits^{\scriptstyle#1}_{\scriptstyle#2}}}

\def\harrl#1#2{\smash{\mathop{\hbox to .5in{\leftarrowfill}}
\limits^{\scriptstyle#1}_{\scriptstyle#2}}}

\newcommand{\be}{\begin{equation}}
\newcommand{\ee}{\end{equation}}
\newcommand{\bea}{\begin{eqnarray}}
\newcommand{\eea}{\end{eqnarray}}

\begin{document}
\begin{titlepage}
\begin{center}
{\bf \Large{Against Supersymmetry\\}}
\end{center}
\vskip 1.0truecm
\centerline{Dan Radu Grigore
\footnote{e-mail: grigore@theory.nipne.ro}}
\vskip5mm
\centerline{Dept. Theor. Phys., Inst. Atomic Phys.,}
\centerline{Bucharest-M\u agurele, MG 6, Rom\^ania}
\vskip 1cm
\centerline{G\"unter Scharf
\footnote{e-mail: scharf@physik.unizh.ch}}
\vskip5mm
\centerline{Institute of Theor. Phys., University of Z\"urich}
\centerline{Winterthurerstr., 190, Z\"urich CH-8057, Switzerland}
\vskip 2cm
\bigskip \nopagebreak
\begin{abstract}
\noindent
We consider the massless supersymmetric vector multiplet in a purely quantum
framework and propose a power counting formula. Then we prove that the interaction Lagrangian for a massless supersymmetric non-Abelian gauge theory (SUSY-QCD) is uniquely determined by some natural assumptions, as in the case of Yang-Mills models, however we do have anomalies in the second order of perturbation theory. The result can be easily generalized to the case when massive multiplets are present, but one finds out that the massive and the massless Bosons must be decoupled, in contradiction with the standard model. Going to the second order of perturbation theory produces an anomaly which cannot be eliminated. We make a thorough analysis of the model working only with the component fields. 
\end{abstract}

\newpage\setcounter{page}1
\end{titlepage}
\section{Introduction\label{introd}}

A quantum field theory should provide two items: the Hilbert space of the physical states and the (perturbative) expression of the scattering matrix. In perturbation theory the Hilbert space is generated from the vacuum by some set of free fields i.e. it is a Fock space. In theories describing higher spin particles one considers a larger Hilbert space of physical and unphysical degrees of freedom and gives a rule of selection for the physical states; this seems to be the only way of saving unitarity and renormalizability, in the sense of Bogoliubov. In this case one should check that the interaction Lagrangian (i.e. the first order of the $S$-matrix) leaves invariant the physical states. If the preceding picture is available in all detail then one can go very easily to explicit computations of some scattering process. Other constructions of a quantum field theory, as those based on functional integration, are incomplete in our opinion if they are not translated in the operatorial language in such a way that the consistency checks can be easily done.

The construction of the QCD Lagrangian in the causal approach goes as follows 
\cite{Gr1}, \cite{Sc}. The Hilbert space of the massless vector field 
$
v_{\mu}
$
is enlarged to a bigger Hilbert space 
${\cal H}$
including two ghost fields
$u,~\tilde{u}$
which are Fermi scalars of null mass; in
${\cal H}$
we can give a Hermitian structure such that we have
\be
v_{\mu}^{\dagger} = v_{\mu} \qquad u^{\dagger} = u, \qquad \tilde{u}^{\dagger} = - \tilde{u}.
\label{hermite-1}
\ee

Then one introduces the {\it gauge charge} $Q$ according to:
\bea
Q \Omega = 0, \qquad Q^{\dagger} = Q,
\nonumber \\
~[ Q, v_{\mu} ] = i \partial_{\mu}u,
\nonumber \\
~\{ Q, u \} = 0, \qquad 
\{ Q, \tilde{u} \} = - i~\partial^{\mu}v_{\mu};
\label{gh-charge}
\eea
here 
$
\Omega \in {\cal H}
$
is the vacuum state. Because
$
Q^{2} = 0
$
the physical Hilbert space is given by
$
{\cal H}_{phys} = Ker(Q)/Im(Q).
$

The gauge charge is compatible with the following causal (anti)commutation 
relation:
\be
[ v_{\mu}(x), v_{\nu}(y) ] = i~~g_{\mu\nu}~D_{0}(x-y)
\quad
~\{ u(x), \tilde{u}^{\dagger}(y) \} = - i~D_{0}(x-y)
\label{ccr-vector}
\ee
and the other causal (anti)commutators are null; here
$
D_{m}(x-y)
$
is Pauli-Jordan causal distribution of mass 
$m \geq 0$. 
In fact, the first relation together with the definition of the gauge charge, determines uniquely the second relation as it follows from the Jacobi identity
\bea
[ v_{\mu}(x), \{ \tilde{u}(y), Q \} ] + \{ \tilde{u}(y), [ Q, v_{\mu}(x) ] \}
= \{ Q, [ v_{\mu}(x), \tilde{u}(y) ] \} = 0.
\label{YM+CCR}
\eea
We can then assume that all the fields
$v_{\mu}, u, \tilde{u}$
have the canonical dimension equal to $1$ so the gauge charge raises the
canonical dimension by $1$. It is usefull to convince the reader that the gauge structure above gives the right physical Hilbert space. We do this for the one-particle Hilbert space. The generic form of a state 
$
\Psi \in {\cal H}^{(1)} \subset {\cal H}
$
from the one-particle Hilbert subspace is
\be
\Psi = \left[ \int f_{\mu}(x) v^{\mu}(x) + \int g_{1}(x) u(x) + \int g_{2}(x) \tilde{u}(x) \right] \Omega
\ee
with test functions
$
f_{\mu}, g_{1}, g_{2}
$
verifying the wave equation equation. We impose the condition 
$
\Psi \in Ker(Q) \quad \Longleftrightarrow \quad Q\Psi = 0;
$
we obtain 
$
\partial^{\mu}f_{\mu} = 0 \qquad g_{2} = 0
$
i.e. the generic element
$
\Psi \in {\cal H}^{(1)} \cap Ker(Q)
$
is
\be
\Psi = \left[ \int f_{\mu}(x) v^{\mu}(x) + \int g(x) u(x) \right] \Omega
\label{kerQ}
\ee
with $g$ arbitrary and 
$
f_{\mu}
$
constrained by the transversality condition 
$
\partial^{\mu}f_{\mu} = 0;
$
so the elements of
$
{\cal H}^{(1)} \cap Ker(Q)
$
are in one-one correspondence with couples of test functions
$
(f_{\mu}, g)
$
with the transversality condition on the first entry. Now, a generic element
$
\Psi^{\prime} \in {\cal H}^{(1)} \cap Im(Q)
$
has the form 
\be
\Psi^{\prime} = Q\Phi = \left[ - \int \partial^{\mu}f^{\prime}_{\mu}(x) u(x) 
+ \int \partial_{\mu}g^{\prime}(x) v^{\mu}(x) \right] \Omega
\label{imQ}
\ee
so if
$
\Psi \in {\cal H}^{(1)} \cap Ker(Q)
$
is indexed by
$
(f_{\mu}, g)
$
then 
$
\Psi + \Psi^{\prime}
$
is indexed by 
$
(f_{\mu} + \partial_{\mu}g^{\prime}, g - \partial^{\mu}f^{\prime}_{\mu}).
$
If we take 
$
f^{\prime}_{\mu}
$
conveniently we can make 
$
g = 0.
$
We introduce the equivalence relation 
$
f_{\mu}^{(1)} \sim f_{\mu}^{(2)} \quad \Longleftrightarrow 
f_{\mu}^{(1)} - f_{\mu}^{(2)} = \partial_{\mu}g^{\prime}
$
and it follows that the equivalence classes from
$
Ker(Q)/Im(Q)
$ 
are indexed by equivalence classes of wave functions
$
[f_{\mu}];
$
we have obtained the usual one-particle Hilbert space for the photon. The preceding argument can be generalized to multi-particle Hilbert space \cite{Gr0}; the idea comes from Hodge theory and amounts in finding an homotopy operator 
$
\tilde{Q}
$ 
such that the spectum of the ``Laplace" operator
$
\{Q,\tilde{Q}\}
$
can be easily determined. 

By definition quantum chromodynamics assumes that we have $N$ copies
$
v^{\mu}_{j},~u_{j},~\tilde{u}_{j}\quad j = 1,\dots,N
$
verifying the preceding algebra for any 
$j= 1,\dots,N$.

The interaction Lagrangian 
$t(x)$
is some Wick polynomial acting in the total Hilbert space
${\cal H}$
and verifying the conditions: (a) canonical dimension 
$
\omega(t) = 4;
$
(b) null ghost number
$
gh(t) = 0
$
(where by definition we have
$
gh(v^{\mu}_{j}) = 0, \quad gh(u_{j}) = 1 \quad gh(\tilde{u}_{j}) = - 1
$
and the ghost number is supposed to be additive);
(c) Lorentz covariant; (d) gauge invariance in the sense:
\bea
[Q, t(x)]= i\partial_{\mu} t^{\mu}(x)
\label{gauge}
\eea
for some Wick polynomials
$t^{\mu}$
of canonical dimension 
$
\omega(t^{\mu}) = 4
$
and ghost number
$
gh(t^{\mu}) = 1.
$

The gauge invariance condition guarantees that, after spatial integration the 
interaction Lagrangian
$t(x)$
factorizes to the physical Hilbert space
$Ker(Q)/Im(Q)$
in the adiabatic limit, i.e. after integration over $x$; the condition
(\ref{gauge}) is equivalent to the usual condition of (free) current 
conservation. Expressions of the type
\be
d_{Q}b + \partial_{\mu}t^{\mu}
\ee
with
\be
\omega(b) = \omega(t^{\mu}) = 3 \qquad gh(b) = - 1 \quad gh(t^{\mu}) = 0
\ee
are called {\it trivial Lagrangians} because they induce a null interaction after space integration (i.e. the adiabatic limit) on the physical Hilbert space. One can prove that the condition (\ref{gauge}) restricts drastically the possible form of $t$ i.e. every such expression is, up to a trivial Lagrangian, equivalent to
\be
t = f_{jkl} ( :v_{j}^{\mu} v_{k}^{\nu} \partial_{\nu}v_{l\mu}:
- :v_{j}^{\mu} u_{k} \partial_{\mu}\tilde{u}_{l}:)
\label{qcd}
\ee
where the (real) constants
$
f_{jkl}
$
must be completely antisymmetric. In the the rest of the paper we will skip the
Wick ordering notations. Going to the second order of the perturbation theory produces the Jacobi identity. So we see that, starting from some very natural assumptions, we obtain in an unique way the whole structure of Yang-Mills models. We expect the same thing to happen for more complex models as supersymmetric theories.

If we want to generalize to the supersymmetric case we must include all the fields
$
v_{\mu}, u, \tilde{u}
$
in some supersymmetric multiplets. By definition \cite{GS1} a supersymmetric multiplet is a set of Bose and Fermi fields 
$
b_{j}, f_{A}
$
together with the supercharge operators
$
Q_{a}
$
such that the commutator (resp. the anticommutator) of a Bose (resp. Fermi) field with the supercharges is a linear combination of Fermi (resp. Bose) fields; the coefficients of these linear combinations are partial derivative operators. We must also suppose that the supercharges are part of an extension of the Poincar\'e algebra called the supersymmetric algebra; essentially we have (for $N = 1$ supersymmetry):
\be
Q_{a} \Omega = 0, \quad \bar{Q}_{\bar{a}} \Omega = 0 \quad 
\bar{Q}_{\bar{a}} = (Q_{a})^{\dagger}
\label{vac}
\ee
\bea
\{ Q_{a} , Q_{b} \} = 0, \quad
\{ Q_{a} , \bar{Q}_{\bar{b}} \} - 2 \sigma^{\mu}_{a\bar{b}} P_{\mu} = 0
\label{SUSY}
\eea
and
\be
[ Q_{a}, P_{\mu} ] = 0, \quad
U_{A}^{-1} Q_{a} U_{A} = {A_{a}}^{b} Q_{b}.
\label{poincare}
\ee

Here 
$
U_{A}
$
is a unitary representation of the Poincar\'e group and 
$
P_{\mu}
$
are the infinitesimal generators of the space-time translations.
There are not many ways to do this. We will show that for the 
$
v_{\mu}
$
we must use the vector multiplet and for the ghost fields we must use
chiral multiplets. Then we must impose that $t$ is also supersymmetric
invariant. A natural definition is:
\be
[ Q_{a}, t] = d_{Q}s_{a} + \partial_{\mu}t_{a}^{\mu} \qquad
\omega(s_{a}) = 7/2 \quad gh(s_{a}) = - 1 \qquad
\omega(t_{a}^{\mu}) = 7/2 \quad gh(t_{a}^{\mu}) = 0;
\label{susy-inv}
\ee
this means that after space integration (i.e. the adiabatic limit) we
obtain on the physical Hilbert space an expression commuting with the
supercharges.

In the supersymmetric framework one usually makes a supplementary
requirement, namely that the basic supersymmetric multiplets should be
organized in superfields \cite{BK}, \cite{GGRS}, \cite{So}, \cite{Wei} i.e. fields dependent on space-time variables and some auxiliary Grassmann parameters
$
\theta_{a}, \bar{\theta}_{\bar{a}}.
$
It is showed in \cite{GS1} that
there is a canonical map 
$w \mapsto sw \equiv W$ 
mapping a ordinary Wick monomial
$w(x)$
into its supersymmetric extension
\be
W(x,\theta,\bar{\theta}) \equiv 
\exp\left(i\theta^{a} Q_{a} - i\bar{\theta}^{\bar{a}} \bar{Q}_{\bar{a}}\right);
\label{W-expo} 
\ee
in particular this map associates to every field of the model a superfield. Moreover, one postulates that the interaction Lagrangian $t$ should be of the form
\be
t(x) \equiv \int d\theta^{2} d\bar{\theta}^{2} T(x,\theta,\bar{\theta})
\label{t-T}
\ee
for some supersymmetric Wick polynomial $T$. We expect that the
preceding expression is of the form (\ref{qcd}) plus other monomials
where the super-partners appear.

One can hope to have an uniqueness result for the coupling if one finds out a 
supersymmetric generalization of (\ref{gauge}). A natural candidate would be 
the relation:
\be
[Q, T(x,\theta,\bar{\theta}) ] = {\cal D}T(x,\theta,\bar{\theta}) - H.c.
= {\cal D}T + \bar{\cal D}\bar{T}
\label{gauge-susy}
\ee
where
\bea
{\cal D}_{a} \equiv {\partial \over \partial \theta^{a}} 
- i \sigma^{\mu}_{a\bar{b}} \bar{\theta}^{\bar{b}} \partial_{\mu}
\qquad
\bar{\cal D}_{\bar{a}} \equiv 
- {\partial \over \partial \bar{\theta}^{\bar{a}}}
+ i \sigma^{\mu}_{b\bar{a}} \theta^{b} \partial_{\mu}.
\label{calD}
\eea

We have
\bea
( {\cal D}_{a} T)^{\dagger} = \pm \bar{\cal D}_{\bar{a}} T^{\dagger},
\nonumber \\
\{{\cal D}_{a}, {\cal D}_{b} \} = 0, \quad
\{\bar{\cal D}_{\bar{a}}, \bar{\cal D}_{\bar{b}} \} = 0, \quad
\{{\cal D}_{a}, \bar{\cal D}_{\bar{b}} \} = 
-2 i \sigma^{\mu}_{a\bar{b}}~\partial_{\mu}
\label{DD}
\eea
where in the first formula the sign $+ (-)$ corresponds to a super-Bose (-Fermi)
field. The last relations is used to eliminate space-time divergences
$\partial_{\mu} T^{\mu}(x,\theta,\bar{\theta})$
in the right-hand side of the relation (\ref{gauge-susy}). It is clear that 
(\ref{gauge-susy}) implies (\ref{gauge}).

The most elementary and general way of analyzing supersymmetries is
to work in components and to see later if the solution can be expressed
in terms of superfields. We will prove that in the massless case there
is an unique solution for SUSY-QCD if we consider a weaker form of
(\ref{susy-inv}) i.e. we require that this relation is true only on
physical states:
\be
<\Psi_{1}, ([ Q_{a}, t] - d_{Q}s_{a} - \partial_{\mu}t_{a}^{\mu} ) \Psi_{2} > = 0
\label{susy-inv-phys}
\ee
where 
$
\Psi_{1}, \Psi_{2}  \in Ker(Q)~{\rm modulo}~Im(Q).
$
We remark that (\ref{susy-inv}) implies (\ref{susy-inv-phys}) but not the other way round.

In the next Section we give the structure of the multiplets of the model and we present the gauge structure. In Section \ref{qcd-interaction} we determine the most general form of the
interaction Lagrangian compatible with gauge invariance and prove that we have supersymmetric invariance also. The expression for the ghost coupling seems to be new in the literature. The details of the computation are given in the Appendix.

We also investigate in what sense one can rephrase the result using superfields. An immediate consequence of the analysis in terms of component fields is that one cannot impose (\ref{gauge-susy}). However we can establish a contact with traditional literature based on the so-called Wess-Zumino gauge. 
 
In Section \ref{massive} we extend the result to the massive case hoping to obtain
the minimal supersymmetric extension of the standard model. We obtain a curious obstruction, namely the sector of massive gauge fields and the sector of massless gauge fields must decouple; this does not agree with the standard model.

Unfortunately, if we proceed to the second order of perturbation theory, we obtain a supersymmetric contribution to the anomaly which cannot be eliminated by redefinitions of the chronological products. 

In Section \ref{new} we do the same analysis for the new vector multiplet \cite{GS1} working in components also. In conclusion
$
N = 1
$
supersymmetry and gauge invariance do not seem to be compatible in quantum theory.

\section{The Quantum Superfields of the Model\label{superfields}}
\subsection{The Vector Multiplet\label{vector}} 
The vector multiplet is the collection of fields
$
C, \phi, v_{\mu}, d, \chi_{a}, \lambda_{a}
$
where 
$C, $ 
is real scalar, $\phi$ is a complex scalar,
$
v_{\mu}
$
is a real vector and
$
\chi_{a}, \lambda_{a}
$
are spinor fields. We suppose that all these fields are of mass 
$m \geq 0$.
We can group them in the superfield
\be
V= C + \theta\chi + \bar{\theta} \bar{\chi} + \theta^{2}~\phi 
+ \bar{\theta}^{2}~\phi^{\dagger}
+ (\theta \sigma^{\mu} \bar{\theta})~v_{\mu}
+ \theta^{2}~\bar{\theta}\bar{\lambda} 
+ \bar{\theta}^{2}~\theta\lambda
+ \theta^{2} \bar{\theta}^{2}~ d.
\label{V}
\ee

It is convenient to define the new field:
\be
\lambda^{\prime}_{a} \equiv \lambda_{a} + {i\over 2} \sigma^{\mu}_{a\bar{b}} 
\partial_{\mu}\bar{\chi}^{\bar{b}} \qquad
d^{\prime} \equiv d - {m^{2} \over 4} C
\ee
and then the action of the supercharges is given by
\bea
i~[Q_{a}, C ] = \chi_{a}
\nonumber \\
~\{ Q_{a}, \chi_{b} \} = 2i~\epsilon_{ab} \phi
\nonumber \\
\{ Q_{a}, \bar{\chi}_{\bar{b}} \} = - i~\sigma^{\mu}_{a\bar{b}}
~( v_{\mu} + i~\partial_{\mu}C )
\nonumber \\
~[Q_{a}, \phi ] = 0
\nonumber \\
i~[Q_{a}, \phi^{\dagger} ] = \lambda^{\prime}_{a} 
- i~\sigma^{\mu}_{a\bar{b}}\partial_{\mu}\bar{\chi}^{\bar{b}}
\nonumber \\
i~[Q_{a}, v^{\mu} ] = \sigma^{\mu}_{a\bar{b}} \bar{\lambda^{\prime}}^{\bar{b}}
- i~\partial^{\mu}\chi_{a}
\nonumber \\
~\{ Q_{a}, \lambda^{\prime}_{b} \} = 2i~\epsilon_{ab}~d^{\prime}
- 2i~\sigma^{\mu\rho}_{ab} \partial_{\mu}v_{\rho}
\nonumber \\
\{ Q_{a}, \bar{\lambda}^{\prime}_{\bar{b}} \} = 0
\nonumber \\
~[Q_{a}, d^{\prime} ] = - {1\over 2} \sigma^{\mu}_{a\bar{b}}
\partial_{\mu}\bar{\lambda^{\prime}}^{\bar{b}}.
\label{susy-action1}
\eea

It is a long but straightforward exercise to verify that the
supersymmetric algebra is valid \cite{GS2}. In \cite{GS2} we have also
determined the generic form of the causal (anti)commutation relations:
\bea
~[ C(x), C(y) ] = - i~c_{1}~D_{m}(x-y)
\nonumber \\
~[ C(x), d(y) ] = - i~c_{2}~D_{m}(x-y)
\nonumber \\
~[ C(x), \phi(y) ] = - i~(c_{4} - i c_{3})~D_{m}(x-y)
\nonumber \\
~[ \phi(x), \phi^{\dagger}(y) ] = 
- i~\left( {m^{2}\over 4}~c_{1} + c_{2}\right)~D_{m}(x-y)
\nonumber \\
~[ \phi(x), d(y) ] = {m^{2}\over 4}~(c_{4} - i c_{3})~D_{m}(x-y)
\nonumber \\
~[ \phi(x), v_{\mu}(y) ] = (c_{3} + i c_{4})~\partial_{\mu}D_{m}(x-y)
\nonumber \\
~[ d(x), d(y) ] = - {im^{4}\over 16}~c_{1}~D_{m}(x-y)
\nonumber \\
~[ v_{\mu}(x), v_{\rho}(y) ] = 
i~c_{1}~\partial_{\mu}\partial_{\rho}~D_{m}(x-y)
+ i~\left({m^{2}\over 2}~c_{1} - 2 c_{2}\right)~g_{\mu\rho}~D_{m}(x-y)
\nonumber \\
\left\{ \chi_{a}(x), \chi_{b}(y) \right\} 
= 2 (c_{4} - i c_{3})~~\epsilon_{ab}~D_{m}(x-y),
\nonumber \\
\left\{ \chi_{a}(x), \bar{\chi}_{\bar{b}}(y) \right\} =
c_{1}~\sigma^{\mu}_{a\bar{b}}~\partial_{\mu}D_{m}(x-y)
\nonumber \\
\left\{ \lambda_{a}(x), \lambda_{b}(y) \right\} 
= - {m^{2}\over 2}(c_{4} - i c_{3})~\epsilon_{ab}~D_{m}(x-y),
\nonumber \\
\left\{ \lambda_{a}(x), \bar{\lambda}_{\bar{b}}(y) \right\} =
{m^{2}\over 4}~c_{1}~\sigma^{\mu}_{a\bar{b}}~\partial_{\mu}D_{m}(x-y)
\nonumber \\
\left\{ \chi_{a}(x), \lambda_{b}(y) \right\} 
= - 2i~c_{2}~\epsilon_{ab}~D_{m}(x-y),
\nonumber \\
\left\{ \chi_{a}(x), \bar{\lambda}_{\bar{b}}(y) \right\} =
- i~(c_{4} + i c_{3})~\sigma^{\mu}_{a\bar{b}}~\partial_{\mu}D_{m}(x-y)
\label{CCR-V}
\eea
and the rest of the (anti)commutators are null; here
$
c_{j}~j = 1,\dots,4
$
are some real coefficients. However, if we want that the commutation relations
for the vector field remain unchanged (\ref{ccr-vector}) then we must
require
$
c_{1} = 0,~c_{2} = - {1\over 2}.
$ 
This choice implies in particular that 
$
v_{\mu}
$
and $\phi$ have the canonical dimensions $1$ so the causal commutator
between them should have the order of singularity $- 2$. But this is compatible
only with the choice
$
c_{3} = c_{4} = 0.
$

In the end we find out that the causal (anti)commutation relations:
\bea
~[ C(x), d^{\prime}(y) ] = {i\over 2}~D_{m}(x-y)
\nonumber \\
~[ d^{\prime}(x), d^{\prime}(y) ] = - {i m^{2}\over 4}~D_{m}(x-y)
\nonumber \\
~[ \phi(x), \phi^{\dagger}(y) ] = {i\over 2}~D_{m}(x-y)
\nonumber \\
~[ v_{\mu}(x), v_{\nu}(y) ] = i~g_{\mu\nu}~D_{m}(x-y)
\nonumber \\
\left\{ \chi_{a}(x), \lambda^{\prime}_{b}(y) \right\} 
= i~~\epsilon_{ab}~D_{0}(x-y),
\nonumber \\
\left\{ \lambda^{\prime}_{a}(x), \bar{\lambda}^{\prime}_{\bar{b}}(y) \right\} 
= \sigma^{\mu}_{a\bar{b}}~\partial_{\mu}D_{0}(x-y) 
\label{CCR-V1}
\eea
and the rest of the (anti)commutators are null. More compactly
\be
[ V(X), V(Y) ] = - {1\over 2}~D_{2}(X;Y) 
\ee
where we are using the notations from \cite{GS2} for the possible causal
super-distributions. The canonical dimension of the component fields are
\be
\omega(v_{\mu}) = 1 \qquad \omega(\lambda^{\prime}) = 3/2 \qquad 
\omega(d) = 2  \qquad \omega(\phi) = 1 \quad
\omega(\chi) = {1\over 2} \quad \omega(\lambda^{\prime}) = {3\over 2}.
\label{can1}
\ee

It is natural to assume that 
\be
\omega(\theta) = - 1/2 \qquad
\omega({\cal D}) = 1/2
\ee
as it is usually done in the literature. In this way one can make sense
of the notion of canonical dimension for the vector superfield; more precisely
we have:
\be
\omega(V) = 0.
\label{CAN1}
\ee

\subsection{The Ghost Chiral Multiplets\label{ghosts}}

We require that the ghost fields are also members of some multiplets of chiral type.
We admit that these ghost multiplets are also massless. 
The generic forms of a chiral ghost and anti-ghost superfields are
\bea
U(x,\theta,\bar{\theta}) = a(x) 
+ 2i~\bar{\theta} \bar{\zeta}(x)
+ i~(\theta \sigma^{\mu} \bar{\theta})~\partial_{\mu}a(x)
+ \bar{\theta}^{2}~g(x)
+ \bar{\theta}^{2}~\theta \sigma^{\mu} \partial_{\mu}\bar{\zeta}(x)
\label{U}
\eea
and respectively
\bea
\tilde{U}(x,\theta,\bar{\theta}) = \tilde{a}(x) 
- 2i~\bar{\theta} \bar{\tilde{\zeta}}(x)
+ i~(\theta \sigma^{\mu} \bar{\theta})~\partial_{\mu}\tilde{a}(x)
+ \bar{\theta}^{2}~\tilde{g}(x)
- \bar{\theta}^{2}~\theta \sigma^{\mu} \partial_{\mu}\bar{\tilde{\zeta}}(x)
\label{tildeU}
\eea
where
$
a,~g,~\tilde{a},~\tilde{g}
$
are Fermi scalar fields and
$
\zeta_{a}, \tilde{\zeta}_{a}
$
are Bose spinor fields. Let us remind that we choose them such that \cite{GS2}
\be
(\zeta_{a})^{\dagger} = \bar{\zeta}_{\bar{a}} \qquad
(\tilde{\zeta}_{a})^{\dagger} = - \bar{\tilde{\zeta}}_{\bar{a}}
\label{hermite-2}
\ee
- see (\ref{hermite-1}); the chirality condition means
\be
{\cal D}_{a}U = 0 \qquad {\cal D}_{a}\tilde{U} = 0.
\ee

The action of the supercharges on these fields is
\bea
\{ Q_{a}, a \} = 0 \qquad
\{ Q_{a}, a^{\dagger} \} = 2\zeta_{a} 
\nonumber \\
\{ Q_{a}, g \} = -2i~\sigma^{\mu}_{a\bar{b}}
~\partial_{\mu}\bar{\zeta}^{\bar{b}}
\qquad
\{ Q_{a}, g^{\dagger} \} = 0
\nonumber \\
~[ Q_{a}, \zeta_{b} ] = \epsilon_{ab} g^{\dagger}
\qquad
i [ Q_{a}, \bar{\zeta}_{\bar{b}} ] = \sigma^{\mu}_{a\bar{b}}~\partial_{\mu}a
\eea
and respectively:
\bea
\{ Q_{a}, \tilde{a} \} = 0 \qquad
\{ Q_{a}, \tilde{a}^{\dagger} \} = 2\tilde{\zeta}_{a} 
\nonumber \\
\{ Q_{a}, \tilde{g} \} = 2i~\sigma^{\mu}_{a\bar{b}}
~\partial_{\mu}\bar{\tilde{\zeta}}^{\bar{b}}
\qquad
\{ Q_{a}, \tilde{g}^{\dagger} \} = 0
\nonumber \\
~[ Q_{a}, \tilde{\zeta}_{b} ] = \epsilon_{ab} \tilde{g}^{\dagger}
\qquad
i [ Q_{a}, \bar{\tilde{\zeta}}_{\bar{b}} ] = - \sigma^{\mu}_{a\bar{b}}~
\partial_{\mu}\tilde{a}.
\eea

It is convenient to work  with  the Hermitian (resp. anti-Hermitian) fields
\bea
u \equiv a + a^{\dagger}  \qquad v \equiv - i (a - a^{\dagger})
\nonumber \\
\tilde{u} \equiv \tilde{a} - \tilde{a}^{\dagger} \qquad 
\tilde{v} \equiv - i (\tilde{a} + \tilde{a}^{\dagger})
\label{h-ghost}
\eea
such that we have
\be
u^{\dagger} = u \qquad v^{\dagger} = v \qquad
\tilde{u}^{\dagger} = - \tilde{u} \quad \tilde{v}^{\dagger} = - \tilde{v}.
\ee

Then we have the following action of the supercharges:
\bea
\{ Q_{a}, u \} = 2 \zeta_{a} \qquad
\{ Q_{a}, v \} = 2 i~\zeta_{a} 
\nonumber \\
\{ Q_{a}, g \} = -2i~\sigma^{\mu}_{a\bar{b}}
~\partial_{\mu}\bar{\zeta}^{\bar{b}}
\qquad
\{ Q_{a}, g^{\dagger} \} = 0
\nonumber \\
~[ Q_{a}, \zeta_{b} ] = \epsilon_{ab} g^{\dagger}
\qquad
i [ Q_{a}, \bar{\zeta}_{\bar{b}} ] 
= {1\over 2} \sigma^{\mu}_{a\bar{b}}~\partial_{\mu}(u + i v)
\label{susy-gh}
\eea
and respectively:
\bea
\{ Q_{a}, \tilde{u} \} = - 2 \tilde{\zeta}_{a} \qquad
\{ Q_{a}, \tilde{v} \} = - 2 i~\tilde{\zeta}_{a} 
\nonumber \\
\{ Q_{a}, \tilde{g} \} = 2i~\sigma^{\mu}_{a\bar{b}}
~\partial_{\mu}\bar{\tilde{\zeta}}^{\bar{b}}
\qquad
\{ Q_{a}, \tilde{g}^{\dagger} \} = 0
\nonumber \\
~[ Q_{a}, \tilde{\zeta}_{b} ] = \epsilon_{ab} \tilde{g}^{\dagger}
\qquad
i [ Q_{a}, \bar{\tilde{\zeta}}_{\bar{b}} ] = - {1\over 2} \sigma^{\mu}_{a\bar{b}}~
\partial_{\mu}(\tilde{u} +i \tilde{v}).
\label{susy-anti-gh}
\eea

These relations are consistent with the following canonical dimensions for the fields:
\bea
\omega(u) = 1 \quad \omega(v) = 1 \quad \omega(g) = 2 \quad
\omega(\zeta) = {3\over 2}
\nonumber \\
\omega(\tilde{u}) = 1 \quad \omega(\tilde{v}) = 1 \quad \omega(\tilde{g}) = 2 \quad 
\omega(\tilde{\zeta}) = {3\over 2}.
\label{can2}
\eea

One can define the so-called $R$ symmetry by:
\be
R\Omega = 0 \qquad R^{\dagger} = R
\ee
and
\bea
~[R, C] = 0 \quad 
[ R, \phi] = - 2 \phi \quad [R, \phi^{\dagger}] = 2 \phi^{\dagger} \quad [R, d] = 0
\nonumber \\
~[R, \chi ] = - \chi \quad 
[R, \bar{\chi} ] = \bar{\chi} \quad
[ R, \lambda^{\prime} ] = \lambda^{\prime} \quad
[ R, \bar{\lambda}^{\prime} ] = - \bar{\lambda}^{\prime}
\nonumber \\
~[R, u ] = 0 \quad
[R, v ] = 0 \quad
[R, g ] = - 2 g \quad
[R, g^{\dagger} ] = 2 g^{\dagger} \qquad
[R, \zeta_{a} ] = - \zeta \qquad 
[ R, \bar{\zeta}_{\bar{a}} ] = \bar{\zeta}
\nonumber \\
~[R, \tilde{u} ] = 0 \quad
[R, \tilde{v} ] = 0 \quad
[R, \tilde{g} ] = 0 \quad
[R, \tilde{g}^{\dagger} \} = 0
\nonumber \\
~[R, \tilde{\zeta}_{a} ] = - \tilde{\zeta}_{a} \qquad
[ R, \bar{\tilde{\zeta}}_{\bar{a}} ] = \bar{\tilde{\zeta}}_{\bar{a}}.
\label{R}
\eea

One usually imposes this invariance on the interaction Lagrangian.

\subsection{The Gauge Charge\label{gauge-charge}}

The purpose is to generalize the construction from the Introduction and
to extend naturally the formulas (\ref{gh-charge}) to the supersymmetric
case. We define the gauge charge $Q$ postulating the following properties:
\begin{itemize}
\item
We have
\be
Q \Omega = 0, \qquad Q^{\dagger} = Q.
\label{gauge1}
\ee
\item
The (anti)commutator of $Q$ with a Bose (resp. Fermi) field is a linear combination of Fermi (resp. Bose) fields; the coefficients of this linear combinations are partial differential operators.
\item
The (anti)commutator of $Q$ with a field raises the canonical dimension by an unit;
\item
The (anti)commutator of $Q$ with a field raises the ghost number by an unit;
\item
The gauge charge commutes with the action of the Poincar\'e group; in this way the Poincar\'e group induces an action on the physical space
$
{\cal H}_{\rm phys} \equiv Ker(Q)/Im(Q).
$
\item 
The gauge charge anticommutates with the supercharges:
\be
\{Q,Q_{a}\}= 0;
\ee
in this way the supersymmetric algebra induces an action on the physical space
$
{\cal H}_{\rm phys}.
$
\item
The gauge charges squares to zero as in the Yang-Mills case
\be
Q^{2} = 0.
\ee
\end{itemize}

If one makes the most general ansatz for the action of $Q$ compatible with the preceding conditions one finds out an important result: making some convenient rescaling of the fields, the gauge charge is uniquely determineded preceding assumptions. In the massless case we have
$
d^{\prime}= d
$
and $Q$ is uniquely determined by:
\be
Q \Omega = 0 \qquad Q^{\dagger} = Q
\ee
and
\bea
~[Q, C] = i~v \quad [Q, v^{\mu}] = i \partial^{\mu}u \quad 
[ Q, \phi] = - g^{\dagger} \quad [Q, \phi^{\dagger}] = g \quad [Q, d] = 0
\nonumber \\
\{Q, \chi \} = 2 i \zeta \quad 
\{Q, \bar{\chi} \} = - 2 i \bar{\zeta} \quad
\{Q, \lambda^{\prime} \} = 0
\nonumber \\
\{Q, u \} = 0 \quad
\{Q, v \} = 0 \quad
\{Q, g \} = 0 \quad
\{Q, g^{\dagger} \} = 0 \qquad
[Q, \zeta_{a} ] = 0 \qquad 
[ Q, \bar{\zeta}_{\bar{a}} ] = 0
\nonumber \\
\{Q, \tilde{u} \} = - i~\partial_{\mu}v^{\mu}\quad
\{Q, \tilde{v} \} = 2 i~d \quad
\{Q, \tilde{g} \} = 0 \quad
\{Q, \tilde{g}^{\dagger} \} = 0
\nonumber \\
~[Q, \tilde{\zeta}_{a} ] = - {1\over 2} \sigma^{\mu}_{a\bar{b}} 
\partial_{\mu}\bar{\lambda}^{\prime\bar{b}}
\qquad
[ Q, \bar{\tilde{\zeta}}_{\bar{a}} ] = - {1\over 2} \sigma^{\mu}_{b\bar{a}} \partial_{\mu}{\lambda^{\prime}}^{b}.
\label{gauge4}
\eea

One can express everything in terms of superfields also \cite{GGRS}:
\bea
~[ Q, V ] = U - U^{\dagger} \qquad 
\{Q, U \} = 0
\nonumber \\
\{Q, \tilde{U} \} = - {1\over 16}~{\cal D}^{2}~\bar{\cal D}^{2}V
\eea

As in the Introduction we postulate that the physical Hilbert space is the
factor space
$
{\cal H}_{phys} = Ker(Q)/Im(Q).
$
Using this gauge structure it is easy to prove that the one-particle Hilbert
subspace contains the following type of particles: a) a particle of null mass and helicity $1$ (the photon); b) a particle of null mass and helicity $1/2$ (the photino); c) the ghost states generated by the fields
$
\tilde{g}
$
from the vacuum. These states must be eliminated by imposing the supplementary condition that the physical states have null ghost number. Only the transversal degrees of freedom of
$v_{\mu}$
and
$\lambda^{\prime}_{a}$
are producing physical states.

We can now determine the causal (anti)commutation relations for the
ghost fields. As in the Yang-Mills case - see relation (\ref{YM+CCR}) - one uses the Jacobi identities
\bea
[ b(x), \{ f(y), Q \} ] + \{ f(y), [ Q, b(x) ] \} = \{ Q, [ b(x), f(y) ] \} = 0
\label{susy+CCR}
\eea
where
$
b = C, \phi, \phi^{\dagger}, v_{\mu}, d^{\prime}, \chi, \bar{\chi}, \lambda^{\prime},\bar{\lambda}^{\prime}
$
and
$
f = \tilde{u}, \tilde{g}, \tilde{g}^{\dagger}, \tilde{\zeta}, \bar{\tilde{\zeta}};
$
if we take into account the particular choice we have made for the causal (anti)commutation relations of the vector multiplet one finds out \cite{GS2} that we have
\be
~\{ a(x), \tilde{a}^{\dagger}(y) \} = {i\over 2}~D_{0}(x-y) \qquad
~[ \zeta_{a}(x), \bar{\tilde{\zeta}}_{\bar{b}}(y) ] = - {1\over 4}
\sigma^{\mu}_{a\bar{b}}~\partial_{\mu}D_{0}(x-y)
\label{CCR-chiral-ghost}
\ee
and all other causal (anti)commutators are null. So, like for ordinary
Yang-Mills models, the causal (anti)commutators of the ghost fields are
determined by the corresponding relations of the vector multiplet
fields. 

Finally we mention that we can impose that the vector field is part of the so-called rotor multiplet  \cite{GS2}. The fields of this are 
$
(d,v_{\mu},\lambda_{a})
$
and the supercharges defined through
\bea
~[Q_{a}, d ] = \sigma^{\mu}_{a\bar{b}}~\partial_{\mu}\bar{\lambda}^{\bar{b}}.
\nonumber \\ 
~\{ Q_{a}, \lambda_{b} \} = - i~( \epsilon_{ab} d + \sigma^{\mu\nu}_{ab} F_{\mu\nu})
\nonumber \\
\{ Q_{a}, \bar{\lambda}_{\bar{b}} \} = 0
\nonumber \\
i~[Q_{a}, v^{\mu} ] = \sigma^{\mu}_{a\bar{b}} \bar{\lambda}^{\bar{b}}.
\label{rotor1}
\eea
where
$
F_{\mu\nu} \equiv \partial_{\mu}v_{\nu} - \partial_{\nu}v_{\mu}.
$

However in this case the general form of the gauge charge 
\be
~[Q, v^{\mu}] = i \partial^{\mu}u \quad  [Q, d] = 0 \quad
\{Q, \lambda^{\prime}_{a} \} = \alpha~\sigma^{\mu}_{a\bar{b}}~\partial_{\mu}\bar{\zeta}^{\bar{b}}
\ee
is not compatible with the relation 
$
\{Q, Q_{a}\} = 0.
$
Also if we consider that the ghost multiplets are Wess-Zumino we obtain a contradiction.

\section{Supersymmetric QCD\label{qcd-interaction}}

\subsection{Supersymmetric QCD in Terms of Component Fields}

It is better to illustrate the method we use to find the most general
gauge invariant Lagrangian on the simplest case, namely ordinary QCD,
i.e. we will briefly show how to obtain the expression (\ref{qcd})
as the unique possibility. So, we consider a Wick polynomial $t$ which
is tri-linear in the fields
$
v_{j}^{\mu}, u_{j}, \tilde{u}_{j}
$
has canonical dimension $4$ and null ghost number, is Lorentz covariant and gauge invariant in the sense (\ref{gauge}). First we list all possible monomials compatible with all these requirements; they are:
\be
f^{(1)} = f^{(1)}_{jkl} v_{j}^{\mu} v_{k}^{\nu} \partial_{\mu}v_{l\mu} \qquad
f^{(2)} = f^{(2)}_{jkl} v_{j}^{\mu} v_{k\mu} \partial_{\nu}v_{l}^{\nu}
\ee
and
\be
g^{(1)} = g^{(1)}_{jkl} v^{\mu}_{j} u_{k} \partial_{\mu}\tilde{u}_{l} \qquad
g^{(2)} = g^{(2)}_{jkl} \partial_{\mu}v^{\mu}_{j} u_{k} \tilde{u}_{l} \qquad
g^{(3)} = g^{(3)}_{jkl} v^{\mu}_{j} \partial_{\mu}u_{k} \tilde{u}_{l}.
\ee 

We now list the possible trivial Lagrangians. They are total divergences of null ghost number
\bea
t^{(1)}_{\mu} = t^{(1)}_{jkl} v_{j}^{\nu} v_{k\nu} v_{l\mu} \qquad 
t^{(1)}_{jkl} = t^{(1)}_{kjl}
\nonumber \\
t^{(2)}_{\mu} = t^{(2)}_{jkl} v_{j\mu} u_{k} \tilde{u}_{l}
\eea 
and the co-boundary terms of ghost number $- 1$:
\bea
b^{(1)} = b^{(1)}_{jkl} v_{j}^{\mu} v_{k\mu} \tilde{u}_{l} \qquad 
b^{(1)}_{jkl} = b^{(1)}_{kjl}
\nonumber \\
b^{(2)} = b^{(2)}_{jkl} u_{j} \tilde{u}_{k} \tilde{u}_{l} \qquad 
b^{(2)}_{jkl} = - b^{(2)}_{jlk}.
\eea

Now we proceed as follows: using
$
\partial^{\mu}t^{(1)}_{\mu}
$
it is possible to make
\be
f^{(1)}_{jkl} = - f^{(1)}_{lkj};
\label{A1}
\ee
using 
$d_{Q}b^{(1)}$ 
we can make
\be
f^{(2)}_{jkl} = 0;
\ee
using
$
\partial^{\mu}t^{(2)}_{\mu}
$
it is possible to take
\be
g^{(3)}_{jkl} = 0;
\ee
finally, using 
$d_{Q}b^{(2)}$ 
we can make
\be
g^{(2)}_{jkl} = g^{(2)}_{kjl}.
\label{S1}
\ee

So we are left only with three terms. If we compute 
$
d_{Q}t
$
and use the known identity:
\be
\partial^{2} f_{j} = 0, \quad j = 1,2,3 
\quad \Longrightarrow
(\partial^{\mu}f_{1}) (\partial_{\mu}f_{2}) f_{3} =
{1\over 2} \partial_{\mu} \Bigl[ (\partial^{\mu}f_{1}) f_{2} f_{3}
+ f_{1} (\partial^{\mu}f_{2}) f_{3} - f_{1} f_{2} (\partial^{\mu}f_{3}) \Bigl]
\label{magic}
\ee
the result is
\be
d_{Q}t = i u_{j} A_{j}  + {\rm total~div}
\ee
where:
\bea
A_{j} = - 2 f^{(1)}_{jkl}~\partial^{\nu}v^{\mu}_{k}~\partial_{\mu}v_{l\nu}
+ (f^{(1)}_{lkj} + g^{(2)}_{kjl})~
\partial_{\mu}v^{\mu}_{k}~\partial_{\nu}v_{l}^{\nu}
\nonumber \\
+ (- f^{(1)}_{jkl} + f^{(1)}_{lkj} + f^{(1)}_{klj} + g^{(1)}_{kjl})~
v^{\mu}_{k}~\partial_{\mu}\partial_{\nu}v^{\nu}_{l}.
\eea

Now the gauge invariance condition (\ref{gauge}) becomes
\be
u_{j} A_{j} = \partial_{\mu}t^{\mu}.
\ee

From power counting arguments it follows that the general form for 
$
t^{\mu}
$
is
\be
t^{\mu} = u_{j} t^{\mu}_{j} + (\partial_{\nu}u_{j}) t^{\mu\nu}_{j}.
\ee
We can prove that
$
t^{\mu\nu}_{j} = g^{\mu\nu}~t_{j}
$
from where
$
A_{j} = - \partial^{2}t_{j};
$
making a general ansatz for 
$
t_{j}
$
we obtain that we must have in fact 
\be
A_{j} = 0
\ee
i.e. the following system of equations:
\bea
f^{(1)}_{jkl} = - f^{(1)}_{kjl}
\nonumber \\
f^{(1)}_{lkj} + g^{(2)}_{kjl} = 0
\nonumber \\
- f^{(1)}_{jkl} + f^{(1)}_{lkj} + f^{(1)}_{klj} + g^{(1)}_{kjl} = 0;
\label{A2}
\eea
the first equation, together with (\ref{A1}) amounts to the total antisymmetry of the expression
$
f_{jkl} \equiv f^{(1)}_{jkl}.
$
The second equation from the preceding system gives then
$
g^{(2)}_{kjl} = 0.
$
As a result we obtain the (unique) solution:
\be
t^{(1)} = f^{(1)}_{jkl} ( v_{j}^{\mu} v_{k}^{\nu} \partial_{\nu}v_{l\mu}
- v_{j}^{\mu} u_{k} \partial_{\mu}\tilde{u}_{l});
\label{tym}
\ee
it can be easily be proved, using the formula (\ref{magic}) that 
$
d_{Q}t^{(1)}
$
is indeed a total divergence.

The supersymmetric case goes on the same lines only the computational difficulties grow exponentially because now we have many more terms of the type 
$
f,~g,~t_{\mu},~b.
$

We would expect to obtain the expression (\ref{tym}) together with terms with supersymmetric partners. The details are given in the Appendix. It is remarkable that we obtain only two possible solutions namely the usual Yang-Mills solution for QCD~
$t^{(1)}$ 
given above by (\ref{tym}) and
\bea
t^{(2)} = f^{\prime}_{jkl} [ 
(\lambda^{\prime}_{j} \sigma_{\mu} \bar{\lambda}^{\prime}_{k}) v_{l}^{\mu}
+ 2 i~(\lambda^{\prime}_{j} \tilde{\zeta}_{k}) u_{l}
+ 2 i~(\bar{\lambda}^{\prime}_{k} \bar{\tilde{\zeta}}_{j}) u_{l} ].
\label{tsusy}
\eea

We impose now supersymmetric invariance condition (\ref{susy-inv}) on 
\be
t \equiv t^{(1)} + t^{(2)}
\label{t12}
\ee
and we hope to obtain a non-trivial solution. If we are successful we
must also go the the second order of perturbation theory. First we
consider only the terms bilinear in
$v_{\mu}$
and linear in
$\lambda^{\prime}$
from the commutator 
$
[Q_{a}, t]
$
and impose that condition that they are a sum of a co-boundary and a
total divergence. It is not very hard to prove that we obtain the
restriction 
$
f^{(2)}_{jkl} = - i~f^{(1)}_{jkl}
$
i.e. the interaction Lagrangian should be:
\be
t = f_{jkl} [ v_{j}^{\mu} v_{k}^{\nu} \partial_{\nu}v_{l\mu}
- i (\lambda^{\prime}_{j} \sigma_{\mu} \bar{\lambda}^{\prime}_{k}) v_{l}^{\mu}
- v_{j}^{\mu} u_{k} \partial_{\mu}\tilde{u}_{l}
+ 2 ~(\lambda^{\prime}_{j} \tilde{\zeta}_{k}) u_{l}
+ 2 ~(\bar{\lambda}^{\prime}_{k} \bar{\tilde{\zeta}}_{j}) u_{l} ]
\label{susy-t}
\ee
where we have simplified the notation:
$
f_{jkl} \equiv f^{(1)}_{jkl}.
$
We note that this interaction Lagrangian is $R$-invariant. The first two terms are standard in the literature - see for instance \cite{Ro1}, \cite{Ro2}. The next contribution is the standard ghost coupling from the Yang-Mills theory \cite{Gr0}, \cite{Sc}. The last two contributions to the ghost coupling in seems to be absent from this analysis. So, gauge invariance in the first order is not true for the expression from \cite{Ro2}.

We can prove after some computation that the preceding expression verifies the following equation:
\be
[ Q_{a}, t] = d_{Q}s_{a} + \partial^{\mu}t_{\mu a} 
- 2~f_{jkl} ( v_{j}^{\mu} u_{k} \partial_{\mu}\tilde{\zeta}_{la}
+ 2 i \sigma^{\mu\nu}_{ab} v_{j\nu} \partial_{\mu} \tilde{\zeta}_{k}^{b} u_{l})
\label{brocken-susy-inv}
\ee
where
\bea
s_{a} \equiv f_{jkl} [ - 2 \sigma^{\mu\nu}_{ab} v_{j\mu} v_{k\nu} \tilde{\zeta}_{l}^{b}
- i \chi_{ja} v_{k}^{\mu} \partial_{\mu}\tilde{u}_{l}
- 2 i (\lambda^{\prime}_{j}\tilde{\zeta}_{k}) \zeta_{la} 
- 2 i (\bar{\lambda}^{\prime}_{k}\bar{\tilde{\zeta}}_{j}) \zeta_{la}
\nonumber \\
+ i \tilde{v}_{j}\sigma^{\mu}_{a\bar{b}} \bar{\lambda}^{\prime \bar{b}}_{k} v_{l\mu}
- 2 \tilde{v}_{j} \tilde{\zeta}_{ka} u_{l}
- 2 \lambda^{\prime}_{j} \phi_{k} u_{l} ]
\label{sa}
\eea
and
\bea
t_{\mu a} \equiv  f_{jkl} \Bigl[ 
\partial_{\mu}\chi_{ja} u_{k} \tilde{u}_{l} 
-  \chi_{ja} \partial_{\mu}u_{k} \tilde{u}_{l}
+ \chi_{ja} v_{k}^{\nu} \partial_{\mu}v_{l\nu}
- {i\over 2} \epsilon_{\mu\nu\rho\alpha} \sigma^{\alpha}_{a\bar{b}}
\bar{\lambda}^{\prime \bar{b}}_{j} v_{k}^{\nu} v_{l}^{\rho}
\nonumber \\
- \chi_{ja} v_{k}^{\nu} \partial_{\nu}v_{l\mu}
+ i (\lambda_{j}^{\prime} \sigma_{\mu} \bar{\lambda}_{k}^{\prime}) \chi_{la}
+ g_{\mu\nu} \sigma^{\nu}_{a\bar{b}} \tilde{v}_{j} \bar{\lambda}_{k}^{\prime \bar{b}} u_{l}
+ 4 i g_{\mu\nu} \sigma^{\nu\rho}_{ab} v_{j\rho} \tilde{\zeta}_{k}^{b} u_{l} \Bigl].
\eea

So, its seems that the last two terms from (\ref{brocken-susy-inv}) - which cannot be rewritten as 
$
d_{Q}s_{a} + {\rm total~divergence}
$
- are apparently spoiling the supersymmetric invariance condition (\ref{susy-inv}). 

However, as we have said in the Introduction, there is a natural way to save supersymmetric invariance namely
to impose instead of (\ref{susy-inv}) the weaker form (\ref{susy-inv-phys})
\bea
<\Psi_{1}, ([ Q_{a}, t ] - d_{Q}s_{a} - \partial_{\mu}t_{a}^{\mu} ) \Psi_{2}> = 0
\nonumber
\eea
with 
$
\Psi_{1}, \Psi_{2} \in Ker(Q)~{\rm modulo}~Im(Q).
$
We first take 
$
\Psi
$
to be generated from the vacuum by the physical fields
$
v_{j}^{\mu},~\lambda^{\prime}_{j},~d^{\prime}_{j};
$ 
if we consider as before, the terms bilinear in
$v_{\mu}$
and linear in
$\lambda^{\prime}$
from the commutator 
$
[Q_{a}, t]
$
we obtain the interaction Lagrangian should be given by (\ref{susy-t}). It follows that we also have (\ref{brocken-susy-inv}); however, if we apply this relation on a physical states
$
\Psi_{j}
$
we obtain that (\ref{susy-inv-phys}) is true; indeed the extra terms give zero in this average.  If we substitute in (\ref{susy-inv-phys}) 
$
\Psi_{j} \rightarrow \Psi_{j} + Q \Phi_{j}
$
the relation stays true (one has to use the anticommutativity of $Q$ with the supercharges.)
So we have obtained an unique solution for supersymmetric QCD as in the Yang-Mills case.

\subsection{Supersymmetric QCD in Terms of Superfields}

From the preceding Subsection we conclude that we cannot express the interaction Lagrangian in terms of superfields such that (\ref{t-T}) and (\ref{gauge-susy}) are true; indeed if this would be true then we would have (\ref{susy-inv}) with 
$
s_{a} = 0.
$
Even the possibility of obtaining the expression (\ref{susy-t}) in the form (\ref{t-T}) is very unlikely. Nevertheless let us start from the  superfield expression of the interaction suggested by the classical analysis \cite{GS2}. 

Arguments from classical field theory suggests that the interaction should be an expression of the type
\be
t_{\rm classical} = t_{\rm classical}^{(1)} + t_{\rm classical}^{(2)}
\label{t-class}
\ee
where
\bea
t_{\rm classical}^{(1)} \equiv - {i\over 8} \int d\theta^{2} d\bar{\theta}^{2}~ 
f_{jkl}~[ V_{j} {\cal D}^{a}V_{k}~\bar{\cal D}^{2}{\cal D}_{a}V_{l} - H.c. ]
\nonumber \\
t_{\rm classical}^{(2)} \equiv 2 i~\int d\theta^{2} d\bar{\theta}^{2}~f_{jkl}~
V_{j} (U_{k} + U_{k}^{\dagger})~(\tilde{U}_{l} + \tilde{U}_{l}^{\dagger}) 
\eea
(see \cite{GS2}). After a tedious computation one obtains up to a total divergence
\bea
t_{\rm classical}^{(1)} =  f_{jkl}~\Bigl[ 
C_{j} (\partial_{\nu}\chi_{k} \sigma^{\mu\nu} \partial_{\mu} \lambda^{\prime}_{l})
- C_{j} (\partial_{\nu}\bar{\chi}_{k} \sigma^{\mu\nu} \partial_{\mu} \bar{\lambda}^{\prime}_{l})
- 2~\partial_{\mu}C_{j} v_{k}^{\mu} d_{l}
\nonumber \\
+ {1\over 2}~(\chi_{j} \sigma^{\mu} \partial_{\mu} \bar{\chi}_{k})~d_{l}
- {1\over 2}~d_{j} (\partial_{\mu}\chi_{j} \sigma^{\mu} \bar{\chi}_{k})~d_{l}
- 4 i~\phi_{j} \phi_{k}^{\dagger} d_{l}
\nonumber \\
+ v_{j\mu}v_{k\nu}\partial^{\nu}v_{l}^{\mu}
- i~(\lambda^{\prime}_{j} \sigma_{\mu} \bar{\lambda}^{\prime}_{k}) v_{l}^{\mu}
\nonumber \\
- i~(\chi_{j} \sigma_{\mu\nu} \partial^{\mu}\lambda^{\prime}_{k}) v_{j}^{\nu}
- i~(\bar{\chi}_{j} \bar{\sigma}_{\mu\nu} \partial^{\mu}\bar{\lambda}^{\prime}_{k}) v_{j}^{\nu}
+ {1\over 2}~(\chi_{j} \partial_{\mu}\lambda^{\prime}_{k}) v_{l}^{\nu}
- {1\over 2}~(\bar{\chi}_{j} \partial_{\mu}\bar{\lambda}^{\prime}_{k}) v_{l}^{\nu}
\nonumber\\
+ 2 i~(\chi_{j} \lambda^{\prime}_{k})d_{l}
- 2 i~(\bar{\chi}_{j} \bar{\lambda}^{\prime}_{k})d_{l} \Bigl] 
+ {\rm total~divergence};
\label{t-class-1}
\eea
the third line is suggested in the literature e.g. \cite{H} see formula
(C.1c) and appears in (\ref{susy-t}) also. The expression of the ghost coupling is much more complicated:
\bea
t_{\rm classical}^{(2)} = f_{jkl}~\Bigl[ - v^{\mu}_{j} u_{k} \partial_{\mu}\tilde{u}_{l}
+ v^{\mu}_{j} \partial_{\mu}v_{k} \tilde{v}_{l}
- 2 d_{j} u_{k} \tilde{v}_{l}
- 2~\phi_{j} g_{k} \tilde{v}_{l} - 2~\phi^{\dagger}_{j} g^{\dagger}_{k} \tilde{v}_{l}
\nonumber \\
+ 2 i~\phi_{j} u_{k} \tilde{g}_{l} 
+ 2 i~\phi^{\dagger}_{j} u^{\dagger}_{k} \tilde{g}_{l}
+ 2 i~C_{j} g_{k}^{\dagger} \tilde{g}_{l} 
+ 2 i~C_{j} g_{k} \tilde{g}_{l}^{\dagger}
\nonumber \\
- i~(\chi_{j}\sigma^{\mu}\bar{\zeta}_{k}) \partial_{\mu}\tilde{u}_{l}
+i~(\zeta_{j}\sigma^{\mu}\bar{\chi}_{k}) \partial_{\mu}\tilde{u}_{l}
+ 2 i~(\lambda^{\prime}_{j}\zeta_{k}) \tilde{v}_{l} 
+ 2 i~(\bar{\lambda}^{\prime}_{j}\bar{\zeta}_{k}) \tilde{v}_{l}
\nonumber \\
- (\partial_{\mu}\chi_{j}\sigma^{\mu}\bar{\zeta}_{k}) \tilde{v}_{l}
+ (\chi_{j}\sigma^{\mu}\partial_{\mu}\bar{\zeta}_{k}) \tilde{v}_{l}
+ (\partial_{\mu}\zeta_{j}\sigma^{\mu}\bar{\chi}_{k}) \tilde{v}_{l}
- (\zeta_{j}\sigma^{\mu}\partial_{\mu}\bar{\chi}_{k}) \tilde{v}_{l}
\nonumber \\
+ 2~(\lambda^{\prime}_{j}\tilde{\zeta}_{k}) u_{l}
+ 2~(\bar{\lambda}^{\prime}_{j}\bar{\tilde{\zeta}}_{l}) u_{l}
+ i~(\partial_{\mu}\chi_{j}\sigma^{\mu}\bar{\tilde{\zeta}}_{l}) u_{k}
- i~(\chi_{j}\sigma^{\mu}\partial_{\mu}\bar{\tilde{\zeta}}_{l}) u_{k}
\nonumber \\
+ i~(\partial_{\mu}\tilde{\zeta}_{l}\sigma^{\mu}\bar{\chi}_{j}) u_{k}
- i~(\tilde{\zeta}_{l}\sigma^{\mu}\partial_{\mu}\bar{\chi}_{j}) u_{k}
- (\chi_{j}\sigma^{\mu}\bar{\tilde{\zeta}}_{l}) \partial_{\mu}v_{k}
+ (\tilde{\zeta}_{l}\sigma^{\mu}\bar{\chi}_{j}) \partial_{\mu}v_{k}
\nonumber \\
- 4 i~\phi^{\dagger}_{j} (\zeta_{k}\tilde{\zeta}_{l})
+ 4 i~\phi_{j} (\bar{\zeta}_{k}\bar{\tilde{\zeta}}_{l})
- 2~C_{j} (\partial_{\mu}\zeta_{k}\sigma^{\mu}\bar{\tilde{\zeta}}_{l})
+ 2~C_{j} (\zeta_{k}\sigma^{\mu}\partial_{\mu}\bar{\tilde{\zeta}}_{l})
\nonumber \\
- 2~C_{j} (\partial_{\mu}\tilde{\zeta}_{l}\sigma^{\mu}\bar{\zeta}_{k})
+ 2~C_{j} (\tilde{\zeta}_{l}\sigma^{\mu}\partial_{\mu}\bar{\zeta}_{k})
- 2 i~v_{j}^{\mu}~(\zeta_{k}\sigma_{\mu}\bar{\tilde{\zeta}}_{l})
- 2 i~v_{j}^{\mu}~(\tilde{\zeta}_{l}\sigma_{\mu}\bar{\zeta}_{k})
\nonumber \\
+ 2~(\chi_{j}\zeta_{k}) \tilde{g}_{l}
+ 2~(\bar{\chi}_{j}\bar{\zeta}_{k}) \tilde{g}_{l}^{\dagger}
- 2~(\chi_{j}\tilde{\zeta}_{l}) g_{k}
+ 2~(\bar{\chi}_{j}\bar{\tilde{\zeta}}_{l}) g_{k}^{\dagger} \Bigl]
+ {\rm total~divergence}
\label{t-class-2}
\eea

Now one can eliminate trivial terms following the procedure given in the table from the Appendix. As a result the total expression is of the following form:
\be
t_{\rm classical} = t + d_{Q}b + \partial^{\mu}t_{\mu} + \tilde{t}
\ee
where $t$ is given by (\ref{susy-t}) and
\bea
\tilde{t} \equiv f_{jkl} \Bigl[
{i\over 2}~(\partial_{\mu}\chi_{j} \sigma^{\mu} \bar{\chi}_{k}) \partial_{\nu}v^{\nu}_{l}
+ {i\over 2}~(\chi_{j} \sigma^{\nu} \partial_{\nu}\bar{\chi}_{k}) \partial_{\mu}v_{l}^{\mu}
+ \phi_{j}^{\dagger} (\chi_{k} \sigma^{\mu} \partial_{\mu}\bar{\lambda}^{\prime}_{l})
- \phi_{j} (\partial_{\mu}\lambda^{\prime}_{l} \sigma^{\mu} \bar{\chi}_{k})
\nonumber \\
+ 2 i~\phi_{j} u_{k} \tilde{g}_{l}
+ 2 i~\phi^{\dagger}_{j} u_{k} \tilde{g}_{l}^{\dagger}
+ 2 i~C_{j} g_{k}^{\dagger} \tilde{g}_{l}
+ 2 i~C_{j} g_{k} \tilde{g}_{l}^{\dagger}
\nonumber \\
+ i~(\partial_{\mu}\chi_{j}\sigma^{\mu}\bar{\tilde{\zeta}}_{l}) u_{k}
- i~(\chi_{j}\sigma^{\mu}\partial_{\mu}\bar{\tilde{\zeta}}_{l}) u_{k}
+ i~(\partial_{\mu}\tilde{\zeta}_{l}\sigma^{\mu}\bar{\chi}_{k}) u_{k}
- i~(\tilde{\zeta}_{l}\sigma^{\mu}\partial_{\mu}\bar{\chi}_{k}) u_{k}
\nonumber \\
+ (\chi_{j}\sigma^{\mu}\partial_{\mu}\bar{\tilde{\zeta}}_{l}) v_{k}
- (\partial_{\mu}\tilde{\zeta}_{l}\sigma^{\mu}\bar{\chi}_{k}) v_{k}
+ 2~(\chi_{j}\zeta_{k}) \tilde{g}_{l}
+ 2~(\bar{\chi}_{j}\bar{\zeta}_{k}) \tilde{g}_{l}^{\dagger} \Bigl]
\eea
is not trivial. So we do not have an exact match between classical and quantum analysis. However, there is a way of eliminating the last term used in the literature, namely to compute
$
t_{\rm classical}
$
in the so-called Wess-Zumino gauge \cite{WB}. This means that the superfield $V$ can be written as
\be
V = A + A^{\dagger} + V^{\prime}
\ee 
where:
\be
V^{\prime} = (\theta \sigma^{\mu} \bar{\theta})~v_{\mu}
+ \theta^{2}~\bar{\theta}\bar{\lambda}^{\prime} 
+ \bar{\theta}^{2}~\theta\lambda^{\prime}
+ \theta^{2} \bar{\theta}^{2}~ d^{\prime}
\label{V-prime}
\ee
and
\be
A = {1\over 2}~C + \bar{\theta} \bar{\chi} + \bar{\theta}^{2}~\phi^{\dagger} 
+ {i\over 2}~(\theta \sigma^{\mu} \bar{\theta})~\partial_{\mu}C
- {i\over 2}~\bar{\theta}^{2}~(\theta\sigma^{\mu}\partial_{\mu}\bar{\chi}).
\ee

If instead of $V$ one uses $V^{\prime}$ then we have
$
\tilde{t} \rightarrow 0 
$
so we have
\be
t_{\rm classical} \rightarrow t + d_{Q}b + \partial^{\mu}t_{\mu}.
\ee

The conclusion is that our expression for the interaction Lagrangian $t$
can be obtained from (\ref{t-class}) - (\ref{t-class-2}) if we substitute
$
V_{j} \rightarrow V_{j}^{\prime}.
$ 

Gauge invariance and supersymmetric invariance are not very conveniently expressed in terms of 
$
V_{j}^{\prime}
$ 
so one must work with the interaction Lagrangian (\ref{t-class}) in every order of the perturbation theory and make at the end
$
V_{j} \rightarrow V_{j}^{\prime}.
$
However, going to the second order of perturbation theory might be more difficult in this superfield formalism than working in component fields.

\subsection{Second Order of Perturbation Theory}

One can go to the second order of perturbation theory very easily in components and compute the anomaly as in \cite{Gr1}; we have in (\ref{gauge})
\be
t^{\mu} = f_{jkl} \left[ 
{1\over 2} u_{j} v_{k\nu} (\partial^{\nu}v_{l}^{\mu} - \partial^{\mu}v_{l}^{\nu})
- {1\over 2} u_{j} u_{k} \partial^{\mu}\tilde{u}_{l}
- i(\lambda_{j}^{\prime} \sigma^{\mu} \bar{\lambda}_{k}^{\prime}) u_{l} \right].
\ee

From here we obtain
\be
d_{Q}~[ t(x),t(y) ] = 
i {\partial \over \partial x^{\mu}}~[ t^{\mu}(x),t(y)]
+ i {\partial \over \partial y^{\mu}}~[ t(x),t^{\mu}(y)]
\ee

The anomalies are obtained in the process of causal splitting of this identity; we obtain in general
\be
d_{Q} T(t(x),t(y)) = 
i {\partial \over \partial x^{\mu}}~T(t^{\mu}(x),t(y))
+ i {\partial \over \partial y^{\mu}}~T(t(x),t^{\mu}(y)) + A(x,y)
\ee
where
$
T(a(x),b(y))
$
is the chronological product associated to the Wick monomials $a$ and $b$ and
$
A(x,y)
$
is the anomaly which spoils the gauge invariance in the second order. One finds out that
\be
A(x,y) = A^{\rm YM}(x,y) + A^{\rm susy}(x,y)
\ee
where the first term already appears in the pure Yang-Mills case and can be eliminated (as a co-boundary plus a total divergence) if and only if one imposes Jacobi identity on the constants
$
f_{jkl}.
$
The second term from the anomaly is of purely supersymmetric nature:
\be
A^{\rm susy}(x,y) = a^{\rm susy}(x)~\delta(x - y)
\ee
with 
\be
a^{\rm susy} = f_{jkl}~f_{mnl}\left[ - {i \over 2}u_{j} v_{k}^{\mu} 
(\lambda_{m}^{\prime} \sigma_{\mu} \bar{\lambda}_{n}^{\prime})
+ 2 u_{j} u_{k} (\lambda_{m}^{\prime} \tilde{\zeta}_{n}) 
+ 2 u_{j} u_{k} (\bar{\lambda}_{m}^{\prime} \bar{\tilde{\zeta}}_{n}) \right]. 
\label{ano}
\ee

Apparently one cannot write this expression as co-boundary plus a total divergence.


\section{Extension to the Massive Case\label{massive}}

The first thing we must do is to remind how one can give mass to the photon in the causal approach \cite{Gr1} \cite{Sc}. One modifies the framework from the Introduction as follows. 
The Hilbert space of the massive vector field 
$
v_{\mu}
$
is enlarged to a bigger Hilbert space 
${\cal H}$
including three ghost fields
$u,~\tilde{u},~h$.
The first two ones are Fermi scalars and the last is a Bose real scalar field; all the ghost fields are supposed to have the same mass 
$
m > 0
$
as the vector field. In
${\cal H}$
we can give a Hermitian structure such that we have
\be
v_{\mu}^{\dagger} = v_{\mu} \qquad u^{\dagger} = u, \qquad \tilde{u}^{\dagger} = - \tilde{u}
\qquad h^{\dagger} = h
\ee
and we also convene that
$
gh(h) = 0.
$
Then one introduces the {\it gauge charge} $Q$ according to:
\bea
Q \Omega = 0, \qquad Q^{\dagger} = Q,
\nonumber \\
~[ Q, v_{\mu} ] = i \partial_{\mu}u, \qquad
[Q , h ] = i m u
\nonumber \\
~\{ Q, u \} = 0, \qquad 
\{ Q, \tilde{u} \} = - i~(\partial^{\mu}v_{\mu} + m h)
\label{gh-charge-m}
\eea
and, because
$
Q^{2} = 0,
$
the physical Hilbert space is given, as in the massless case, by
$
{\cal H}_{phys} = Ker(Q)/Im(Q).
$

The gauge charge is compatible with the following causal (anti)commutation 
relation:
\be
[ v_{\mu}(x), v_{\nu}(y) ] = i~~g_{\mu\nu}~D_{m}(x-y)
\quad
~\{ u(x), \tilde{u}^{\dagger}(y) \} = - i~D_{m}(x-y)
\quad
[ h(x), h(y) ] = - i~D_{m}(x-y) 
\label{ccr-vector-m}
\ee
and the other causal (anti)commutators are null. We see that the canonical dimension of the scalar ghost field must be 
$
\omega(h) = 1.
$
It is an easy exercise to determine the physical space in the one-particle sector: the equivalence classes are indexed by wave functions of the form 
\be
\Psi_{0} = \int f_{\mu}(x) v^{\mu}(x) \qquad \partial^{\mu}f_{\mu} = 0.
\ee
The general argument can be found in \cite{Gr1}. 

To go to the supersymmetric case we must include the field $h$ into a supersymmetric multiplet. Again, the natural candidate is a chiral field we take:
\bea
B(x,\theta,\bar{\theta}) = b(x) 
+ 2~\bar{\theta} \bar{\psi}(x)
+ i~(\theta \sigma^{\mu} \bar{\theta})~\partial_{\mu}b(x)
+ \bar{\theta}^{2}~f(x)
\nonumber \\
- i~\bar{\theta}^{2}~\theta \sigma^{\mu} \partial_{\mu}\bar{\psi}(x)
+ {m^{2} \over 4}~\theta^{2} \bar{\theta}^{2}~b(x)
\label{chiral}
\eea
where
$b, f$
are complex scalars and
$\psi$
is a spinor field. The chirality conditions is:
\be
{\cal D}_{a}B = 0.
\ee
Some mass-dependent extra-terms 
$
{m^{2} \over 4}~\theta^{2} \bar{\theta}^{2}~u(x)
$
and
$
{m^{2} \over 4}~\theta^{2} \bar{\theta}^{2}~\tilde{u}(x)
$
should be included in the expressions of (\ref{U}) and (\ref{tildeU}) of $U$ and $\tilde{U}$ from subsection \ref{ghosts} respectively. 

The action of the supercharges can be taken to be:
\bea
i~[Q_{a}, b ] = 0 \qquad
i~[Q_{a}, b^{\dagger} ] = 2\psi_{a} 
\nonumber \\
~[Q_{a}, f ] = -2~\sigma^{\mu}_{a\bar{b}}~\partial_{\mu}\bar{\psi}^{\bar{b}}
\qquad
~[Q_{a}, f^{\dagger} ] = 0
\nonumber \\
~\{ Q_{a}, \psi_{b} \} = i~\epsilon_{ab} f^{\dagger}
\qquad
\{ Q_{a}, \bar{\psi}_{\bar{b}} \} = \sigma^{\mu}_{a\bar{b}}~\partial_{\mu}b.
\label{susy-h}
\eea

These relations are compatible with the following canonical dimensions:
\be
\omega(b) = 1 \qquad \omega(\psi) = 3/2 \qquad \omega(f) = 2.
\ee

It is convenient to work with the self-adjoint bosonic ghost fields:
\be
h \equiv b + b^{\dagger} \qquad h^{\prime} \equiv - i (b - b^{\dagger}).
\ee

Next we must find the general form of the gauge charge. We proceed as in Subsection \ref{gh-charge} but we {\bf do not} assume that the (anti)commutator with the gauge charge raises the canonical dimension by an unit; we find out that in the massive case $Q$ is uniquely determined by:
\be
Q\Omega = 0 \qquad Q^{\dagger} = Q
\ee
and
\bea
~[Q, C] = i~v \quad [Q, v^{\mu}] = i \partial^{\mu}u \quad 
[ Q, \phi] = - g^{\dagger} \quad [Q, \phi^{\dagger}] = g \quad [Q, d^{\prime}] = 0
\nonumber \\
\{Q, \chi \} = 2 i \zeta \quad 
\{Q, \bar{\chi} \} = - 2 i \bar{\zeta} \quad
\{Q, \lambda^{\prime} \} = 0
\nonumber \\
\{Q, u \} = 0 \quad
\{Q, v \} = 0 \quad
\{Q, g \} = 0 \quad
\{Q, g^{\dagger} \} = 0 \qquad
[Q, \zeta_{a} ] = 0 \qquad 
[ Q, \bar{\zeta}_{\bar{a}} ] = 0
\nonumber \\
\{Q, \tilde{u} \} = - i~(\partial_{\mu}v^{\mu} + m h)
\nonumber \\
\{Q, \tilde{v} \} = i~( 2d^{\prime} + m h^{\prime} + m^{2} C)
\nonumber \\
\{Q, \tilde{g} \} = - m^{2} \phi^{\dagger} - i m f \quad
\{Q, \tilde{g}^{\dagger} \} = - m^{2} \phi + i m f^{\dagger} 
\nonumber \\
~[Q, \tilde{\zeta}_{a} ] = - {1\over 2} \sigma^{\mu}_{a\bar{b}} 
\partial_{\mu}\bar{\lambda}^{\prime\bar{b}} - m \psi_{a} - {i \over 2} m^{2} \chi_{a}
\qquad 
~[ Q, \bar{\tilde{\zeta}}_{\bar{a}} ] = - {1\over 2} \sigma^{\mu}_{b\bar{a}} \partial_{\mu}{\lambda^{\prime}}^{b} - m \bar{\psi}_{\bar{a}} + {i \over 2} m^{2} \bar{\chi}_{\bar{a}}
\nonumber \\
~[Q, h ] = i m~u \quad ~[Q, h^{\prime} ] = i m~v \quad
[Q, f ] = - i m g \quad
\{Q,\psi_{a} \} = m \zeta_{a}  \quad
\{Q,\bar{\psi}_{\bar{a}} \} = m \bar{\zeta}_{\bar{a}}. 
\label{gauge5}
\eea

One can express everything in terms of superfields also \cite{GS2}:
\bea
~[ Q, V ] = U - U^{\dagger} \qquad 
\{Q, U \} = 0
\nonumber \\
\{Q, \tilde{U} \} = - {1\over 16}~{\cal D}^{2}~\bar{\cal D}^{2}V - i m B
\qquad 
[ Q, B ] = i m U.
\eea

As in the Introduction we postulate that the physical Hilbert space is the
factor space
$
{\cal H}_{phys} = Ker(Q)/Im(Q).
$
Using this gauge structure it is easy to prove that the one-particle Hilbert
subspace contains the following type of particles: a) a particle of mass $m$ and spin $1$ (the massive photon); b) a particle of mass $m$ and spin $1/2$ (the massive photino); c) a scalar particle of mass $m$.  Only the transversal degrees of freedom of
$v_{\mu}$
and
$\lambda^{\prime}_{a}$
are producing physical states.

We can now determine the causal (anti)commutation relations for the
ghost fields. If we take into account the particular choice we have made
for the causal (anti)commutation relations of the vector multiplet one finds 
out \cite{GS2} that we have
\bea
~\{ a(x), \tilde{a}^{\dagger}(y) \} = {i\over 2}~D_{m}(x-y) \qquad
~[ \zeta_{a}(x), \bar{\tilde{\zeta}}_{\bar{b}}(y) ] = - {1\over 4}
\sigma^{\mu}_{a\bar{b}}~\partial_{\mu}D_{m}(x-y)
\nonumber \\
~[ b(x), b^{\dagger}(y) ] = - {i\over 2}~D_{m}(x-y) \qquad
~\{ \psi_{a}(x), \bar{\tilde{\psi}}_{\bar{b}}(y) \} = - {1\over 4}
\sigma^{\mu}_{a\bar{b}}~\partial_{\mu}D_{m}(x-y)
\label{CCR-chiral-ghost-m}
\eea
and all other causal (anti)commutators are null. So, like for ordinary 
Yang-Mills models, the causal (anti)commutators of the ghost fields are
determined by the corresponding relations of the vector multiplet
fields. So we see that the causal (anti)commutators are uniquely fixed by some
natural requirements. We also can prove that it is not possible to use rotor multiplet instead of the full vector multiplet and it is not possible to use Wess-Zumino multiplets instead of the ghost multiplets.

First, we eliminate terms independent on the fields
$
b, f, \psi_{a},
$
of the type
$
d_{Q}b + \partial_{\mu}t^{\mu}
$
exactly as in the massless case. Now comes an important observation: suppose that we have a Wick monomial 
$
t_{B}
$ 
which has at least one factor 
$
b, f, \psi_{a}
$ 
and it is of canonical dimension 
$
\omega(t_{B}) \leq 4;
$  
if we consider the expression
$
d_{Q}t_{B}
$
then from (\ref{gauge5}) it follows that
$
\omega(d_{Q}t_{B}) \leq 4.
$

This means that the new ($b, f, \psi$-dependent terms) from the interaction
Lagrangian cannot produce terms of canonical dimension $5$ when commuted
with the gauge charge. So, first  we can proceed with the elimination of trivial Lagrangians of canonical dimension $4$ like in the Appendix. This in turn means that the general solution of the gauge invariance problem must be a sum of two expressions of the
type 
$t^{(1)}$ 
and 
$t^{(2)}$ 
from the preceding Section - see (\ref{tym}) and (\ref{tsusy}) - to which one 
must add $b$-dependent terms such that gauge invariance is restored. These 
new terms are easy to obtain. For the Yang-Mills coupling $t$ they are 
well-known \cite{Gr1}
\bea
t^{(1)}_{\rm massive} \equiv
f_{jkl} ( v_{j\mu} v_{k\nu} \partial^{\mu}v_{l}^{\nu}
- v_{j}^{\mu} u_{k} \partial_{\mu}\tilde{u}_{l} )
\nonumber \\
+ f^{\prime}_{jkl} ( h_{j} \partial_{\mu}h_{k} v_{l}^{\mu} 
- m_{k}~h_{j} v_{k\mu} v_{l}^{\mu} 
- m_{k}~h_{j} \tilde{u}_{k} u_{l} )
+ f^{\prime\prime}_{jkl}~h_{j} h_{k} h_{l}
\label{inter}
\eea
and for the second coupling the generic expression is
\bea
t^{(2)}_{\rm massive} \equiv
- i~f_{jkl} [ (\lambda^{\prime}_{j} \sigma_{\mu} \bar{\lambda}^{\prime}_{k}) v_{l}^{\mu}
+ 2 i~(\lambda^{\prime}_{j} \tilde{\zeta}_{k}) u_{l}
+ 2 i~(\bar{\lambda}^{\prime}_{k} \bar{\tilde{\zeta}}_{j}) u_{l}]
\nonumber \\
+ p^{(1)}_{jkl}~(\lambda^{\prime}_{j} \psi_{k}) h_{l} 
+ p^{(2)}_{jkl}~(\bar{\lambda}^{\prime}_{j} \bar{\psi}_{k}) h_{l} 
+ p^{(3)}_{jkl}~(\lambda^{\prime}_{j} \chi_{k}) h_{l} 
+ p^{(4)}_{jkl}~(\bar{\lambda}^{\prime}_{j} \bar{\chi}_{k}) h_{l}. 
\eea

The gauge invariance condition gives well-known constraints on the constants
$
f^{\prime}, f^{\prime\prime}
$
and:
\bea
p^{(3)}_{jkl} = {i \over 2} p^{(4)}_{jkl}~m_{k} \qquad
p^{(4)}_{jkl} = - {i \over 2} p^{(2)}_{jkl}~m_{k} 
\nonumber \\
i~p^{(1)}_{jkl}~m_{l} = 2 f_{jkl}~m_{k} \qquad
i~p^{(2)}_{jkl}~m_{l} = - 2 f_{jkl}~m_{k}.
\eea

We see from the second set of relations that we must have
\be
f_{jkl} = 0 \qquad {\rm for} \qquad m_{k} \not= 0 \quad m_{l} = 0
\ee
which implies that the massive and massless gauge fields must decouple; this
does not agree with the standard model.

Regarding supersymmetric invariance, this means that the argument which
prevents the equation  (\ref{susy-inv}) to be true remains valid but
(\ref{susy-inv-phys}) should stay true.

We note that the anomaly (\ref{ano}) will remain in this case also.

{\bf Remark} The solution found in \cite{GS4} has in fact terms of canonical dimension 5: indeed the third term of formula (4.1) from this paper produces after integration over the Grassmann variables the term 
$
(\psi\zeta) \tilde{g}
$
which is of canonical dimension $5$.

\section{The New Vector Multiplet\label{new}}

The new vector multiplet was introduced in \cite{GS1}; it is the second possibility of a multiplet which contains a vector field. First we give the definition of a Wess-Zumino multiplet: such a multiplet contains a complex scalar field
$\phi$
and a spin $1/2$ Majorana field
$f_{a}$
of the same mass $m$. The supercharges are defined in this case by:
\bea
[Q_{a}, \phi ] = 0, \qquad [\bar{Q}_{\bar{a}}, \phi^{\dagger} ] = 0
\nonumber \\
i~[Q_{a}, \phi^{\dagger} ] = 2 f_{a}, \qquad 
i~[\bar{Q}_{\bar{a}}, \phi] = 2 \bar{f}_{\bar{a}}
\nonumber \\
~\{ Q_{a}, f_{b} \} = - i~m~\epsilon_{ab} \phi, \qquad 
\{ \bar{Q}_{\bar{a}}, \bar{f}_{\bar{b}} \} = i~m~\epsilon_{\bar{a}\bar{b}}
\phi^{\dagger}
\nonumber \\
~\{ Q_{a}, \bar{f}_{\bar{b}} \} = \sigma^{\mu}_{a\bar{b}} \partial_{\mu}\phi,
\qquad 
\{ \bar{Q}_{\bar{a}}, f_{b} \} = \sigma^{\mu}_{b\bar{a}} 
\partial_{\mu}\phi^{\dagger}.
\label{wess}
\eea

The first vanishing commutators are also called (anti) chirality condition.
The causal (anti)commutators are:
\bea
\left[ \phi(x), \phi(y)^{\dagger} \right] = - 2i~ D_{m}(x-y),
\nonumber \\
\left\{ f_{a}(x), f_{b}(y) \right\} = i~~\epsilon_{ab}~m~D_{m}(x-y),
\nonumber \\
\left\{ f_{a}(x), \bar{f}_{\bar{b}}(y) \right\} 
= \sigma^{\mu}_{a\bar{b}}~\partial_{\mu}D_{m}(x-y)
\label{CCR-wess}
\eea
and the other (anti)commutators are zero. 

To construct the new vector multiplet one  and adds a vector index i.e. makes the substitutions
$
\phi \rightarrow v_{\mu} \quad f_{a} \rightarrow \psi_{\mu a};
$
here 
$
v_{\mu}
$
is a complex vector field. In the massless case the action of the supercharges is:
\bea
~[Q_{a}, v_{\mu} ] = 0, \qquad [\bar{Q}_{\bar{a}}, v^{\dagger}_{\mu} ] = 0
\nonumber \\
i~[Q_{a}, v_{\mu}^{\dagger} ] = 2~\psi_{\mu a}, \qquad
i~[\bar{Q}_{\bar{a}}, v_{\mu} ] = 2~\bar{\psi}_{\mu\bar{a}}
\nonumber \\
~\{ Q_{a}, \psi_{\mu b} \} = 0, \qquad
\{ \bar{Q}_{\bar{a}}, \bar{\psi}_{\mu\bar{b}} \} = 0,
\nonumber \\
~\{ Q_{a}, \bar{\psi}_{\mu\bar{b}} \} = \sigma^{\nu}_{a\bar{b}} 
\partial_{\nu}v_{\mu}, \qquad
\{ \bar{Q}_{\bar{a}}, \psi_{\mu b} \} = \sigma^{\nu}_{b\bar{a}} 
\partial_{\nu}v_{\mu}^{\dagger}.
\label{susy-v}
\eea

The gauge structure is done by considering that the ghost and the anti-ghost multiplets are also Wess-Zumino (of null mass) i.e. we can take in the analysis from Subsection \ref{ghosts}
$
g = 0 \quad \tilde{g} = 0.
$

The gauge charge is defined by
\be
Q\Omega = 0 \quad Q^{\dagger} = Q
\ee
and
\bea
~[ Q, v_{\mu} ] = i \partial_{\mu}u, \qquad 
~[ Q, \psi_{\mu a} ] = \partial_{\mu}\zeta_{a}, \qquad 
\nonumber \\
\{ Q, a \} = 0, \qquad
[ Q, \zeta ] = 0
\nonumber \\
\{ Q, \tilde{a} \} = - i~\partial^{\mu}v_{\mu}, \qquad
[ Q, \tilde{\zeta} ] = - \partial_{\mu}\psi^{\mu}_{a}
\eea
and the rest of the (anti)commutators are zero. It is more convenient to work with two real vector fields:
\be
A_{\mu} = v_{\mu} + v_{\mu}^{\dagger} \qquad
B_{\mu} = - i ( v_{\mu} - v_{\mu}^{\dagger})
\ee
such that we have
\bea
~[ Q, A_{\mu} ] = i \partial_{\mu}u \qquad 
[ Q, B_{\mu} ] = i \partial_{\mu}v
\nonumber \\
\{ Q, \tilde{u} \} = - i \partial_{\mu}A^{\mu} \qquad
\{ Q, \tilde{v} \} = - i \partial_{\mu}B^{\mu}. 
\eea

If we have $r$ such multiplets i.e. the fields
$
A_{j}^{\mu}, B_{j}^{\mu}, u_{j}, v_{j}, \tilde{u}_{j}, \tilde{v}_{j} \quad j =1,\dots,r 
$
then it is convenient to define the new fields
$
A_{J}^{\mu}, u_{J}, \tilde{u}_{J}, \quad j =1,\dots,2r 
$
according to
\be
A_{j+r}^{\mu} = B_{j}^{\mu} \quad u_{j+r} = u_{j} \quad \tilde{u}_{j+r} = \tilde{u}_{j}
\qquad j =1,\dots,r. 
\ee

In this way the gauge structure becomes similar to the Yang-Mills case i.e.
\bea
~[ Q, A_{J}^{\mu} ] = i \partial^{\mu}u_{J}
\nonumber \\
\{ Q, u_{J} \} = 0 \qquad \{ Q, \tilde{u}_{J} \} = - i \partial_{\mu}A_{J}^{\mu}
\eea
for 
$
J = 1,\dots,2r.
$

It is not hard to prove that the gauge invariance fixes the interaction Lagrangian to be
of the Yang-Mills type i.e.
\be
t = f_{JKL} ( A_{J}^{\mu} A_{K}^{\nu} \partial_{\nu}A_{L\mu}
- A_{J}^{\mu} u_{K} \partial_{\mu}\tilde{u}_{L} )
\ee
where the constants
$
f_{JKL} 
$
are real (from Hermiticity), completely antisymmetric (from first order gauge invariance) and verify Jacobi identity (from second order gauge invariance); no other terms containing spinor fields are compatible with (first order) gauge invariance. 
One can revert to the old variables defining
\bea
f_{jkl}^{(1)} \equiv f_{jkl} \qquad f_{jkl}^{(2)} \equiv f_{j+r,k+r,l+r}
\nonumber \\
f_{jkl}^{(3)} \equiv f_{j,k+r,l} \qquad f_{jkl}^{(4)} \equiv f_{j+r,k,l+r}.
\eea

Then  $t$ above is a sum of four expressions:
\bea
t^{(1)} = f_{jkl}^{(1)} ( A_{j}^{\mu} A_{k}^{\nu} \partial_{\nu}A_{l\mu}
- A_{j}^{\mu} u_{k} \partial_{\mu}\tilde{u}_{l} ),
\nonumber \\
t^{(2)} = f_{jkl}^{(2)} ( B_{j}^{\mu} B_{k}^{\nu} \partial_{\nu}B_{l\mu}
- B_{j}^{\mu} u_{k} \partial_{\mu}\tilde{u}_{l} ),
\nonumber \\
t^{(3)} = f_{jkl}^{(3)} ( A_{j}^{\mu} B_{k}^{\nu} \partial_{\nu}A_{l\mu} 
- A_{j}^{\mu} A_{k}^{\nu} \partial_{\nu}B_{l\mu}
- B_{j}^{\mu} A_{k}^{\nu} \partial_{\nu}A_{l\mu}
\nonumber \\
- A_{j}^{\mu} u_{k} \partial_{\mu}\tilde{u}_{l} 
+ A_{j}^{\mu} u_{k} \partial_{\mu}\tilde{v}_{l}
+ B_{j}^{\mu} u_{k} \partial_{\mu}\tilde{u}_{l} );
\eea
the expression 
$
t^{(4)}
$
can be obtained from 
$
t^{(3)}
$
performing the change
$
A_{j}^{\mu} \leftrightarrow B_{j}^{\mu}, \quad 
u_{j} \leftrightarrow v_{j}, \quad
\tilde{u}_{j} \leftrightarrow \tilde{v}_{j}.
$

One can show that
$
t^{(3)}
$
is equivalent to a simple expression, namely:
\be
t^{(3)} \sim f_{jkl}^{(3)} ( A_{j}^{\mu} B_{k}^{\nu} \partial_{\nu}A_{l\mu} 
- 2B_{k}^{\mu} A_{j}^{\nu} \partial_{\nu}A_{l\mu}
+ A_{l}^{\mu} \partial_{\mu}u_{k} \tilde{v}_{k} 
- 2 A_{j}^{\mu} v_{k} \partial_{\mu}\tilde{u}_{l} ).
\ee

Now we impose the susy-invariance condition (\ref{susy-inv-phys}); after some computations one obtains the restrictions:
\be
f_{jkl}^{(3)} = i~f_{jkl}^{(1)}, \qquad f_{jkl}^{(2)} = - i~f_{jkl}^{(1)}
\qquad f_{jkl}^{(4)} = - f_{jkl}^{(1)}
\ee 
which, unfortunately, are in contradiction with the reality condition. So, we cannot find a susy-invariant model even in the first order of the perturbation theory.

\section{Conclusions}

There is another possibility, namely to use extended supersymmetries \cite{GS3}. However in this case one finds out immediately that one cannot include the usual ghost fields in some supersymmetric ghost multiplet. Our analysis indicates serious obstacles in constructing supersymmetric extensions of the standard model in the causal approach. 
\newpage
\section{Appendix}
I) We have the following non-trivial possibilities for terms of the type
$
A_{j}A_{k}A_{l}
$
where
$
A_{j},~A_{k},~A_{l} 
$
are fields of the type
$
C, \phi, v^{\mu}, d, \chi, \lambda^{\prime}:
$
\begin{itemize}
\item 
from the sector
$
C\chi\lambda:
$
\be
f^{(1a)}_{jkl} = C_{j} (\partial_{\mu}\chi_{k} \sigma^{\mu\nu} 
\partial_{\nu}\lambda_{l}^{\prime}),
\qquad
f^{(1b)}_{jkl} = C_{j} (\partial_{\mu}\bar{\chi_{k}} \bar{\sigma}^{\mu\nu} \partial_{\nu}\bar{\lambda}_{l}^{\prime})
\ee
and terms with one derivative on 
$C_{j}$
which will be denoted generically by
$
\tilde{F}^{(1)};
$
\item
from the sector 
$
C v_{\mu} v_{\nu}
$
\bea
f^{(2a)}_{jkl} C_{j} \partial^{\mu}v_{k}^{\nu} \partial_{\nu}v_{l\mu},
\qquad
f^{(2b)}_{jkl} C_{j} v_{k}^{\mu} \partial_{\mu}\partial_{\nu}v_{l}^{\nu},
\nonumber \\
f^{(2c)}_{jkl} \epsilon_{\mu\nu\rho\sigma} C_{j} \partial^{\mu}v_{k}^{\nu} \partial^{\rho}v_{l}^{\sigma},
\qquad
f^{(2d)}_{jkl} C_{j} \partial_{\mu}v_{k}^{\mu} \partial_{\nu}v_{l}^{\nu}
\eea
and terms with one derivative on 
$C_{j}$
which will be denoted generically by
$
\tilde{F}^{(2)};
$
\item 
from the sector
$
C v_{\mu} d:
$
\be
f^{(3a)}_{jkl} C_{j} v_{j}^{\mu} \partial_{\mu}d_{l},
\qquad
f^{(3b)}_{jkl} C_{j} \partial_{\mu}v^{\mu}_{k} d_{l}
\ee
and terms with one derivative on 
$C_{j}$
which will be denoted generically by
$
\tilde{F}^{(3)};
$
\item
from the sector
$
C\lambda\lambda:
$
\be
f^{(4a)}_{jkl} C_{j} (\lambda^{\prime}_{k} \sigma^{\mu} \partial_{\mu}\bar{\lambda}^{\prime}_{l}),
\qquad
f^{(4b)}_{jkl} C_{j} (\partial_{\mu}\lambda^{\prime}_{k} \sigma^{\mu} \bar{\lambda}^{\prime}_{l})
\ee
and terms with one derivative on 
$C_{j}$
which will be denoted generically by
$
\tilde{F}^{(4)};
$
\item
from the sector
$
C d d:
$
\be
f^{(5)}_{jkl} C_{j} d_{k} d_{l};
\ee
\item
from the sector
$
\chi\chi \phi:
$
\bea
f^{(6a)}_{jkl} (\partial_{\mu}\chi_{j} \sigma^{\mu\nu} \partial_{\nu}\chi_{k}) \phi_{l},
\qquad
f^{(6b)}_{jkl} (\partial_{\mu}\chi_{j} \sigma^{\mu\nu} \partial_{\nu}\chi_{k}) \phi_{l}^{\dagger},
\nonumber \\
f^{(6c)}_{jkl} (\partial_{\mu}\bar{\chi}_{j} \bar{\sigma}^{\mu\nu} \partial_{\nu}\bar{\chi}_{k}) \phi_{l},
\qquad
f^{(6d)}_{jkl} (\partial_{\mu}\bar{\chi}_{j} \bar{\sigma}^{\mu\nu} \partial_{\nu}\bar{\chi}_{k}) \phi_{l}^{\dagger}
\eea
and terms with one derivative on 
$\phi_{j}$
which will be denoted generically by
$
\tilde{F}^{(6)};
$
\item
from the sector
$
\chi\chi v_{\mu}:
$
\bea
f^{(7a)}_{jkl} (\partial_{\mu}\chi_{j} \sigma^{\mu} \partial_{\nu}\bar{\chi}_{k}) v^{\nu}_{l},
\qquad
f^{(7b)}_{jkl} (\partial_{\mu}\chi_{j} \sigma^{\nu} \partial_{\nu}\bar{\chi}_{k}) v_{l}^{\mu},
\nonumber \\
f^{(7c)}_{jkl} (\partial_{\mu}\partial_{\nu}\chi_{j} {\sigma}^{\mu} \bar{\chi}_{k}) v^{\nu}_{l},
\qquad
f^{(7d)}_{jkl} (\chi_{j} \sigma^{\mu} \partial_{\mu}\partial_{\nu}\bar{\chi}_{k}) v_{l}^{\nu}
\qquad
f^{(7e)}_{jkl} \epsilon_{\mu\nu\rho\lambda} 
(\partial^{\nu}\chi_{j} \sigma^{\rho} \partial^{\lambda}\bar{\chi}_{k}) v_{l}^{\mu}
\eea
and terms with one derivative on 
$v^{\mu}_{l}$
which will be denoted generically by
$
\tilde{F}^{(7)};
$
\item
from the sector
$
\chi\chi d:
$
\be
f^{(8a)}_{jkl} (\partial_{\mu}\chi_{j} \sigma^{\mu} \bar{\chi}_{k}) d_{l},
\qquad
f^{(8b)}_{jkl} (\chi_{j} \sigma^{\mu} \partial_{\mu}\bar{\chi}_{k}) d_{l}
\ee
and terms with one derivative on 
$d_{l}$
which will be denoted generically by
$
\tilde{F}^{(8)};
$
\item
from the sector
$
\chi\lambda \phi:
$
\bea
f^{(9a)}_{jkl} (\partial_{\mu}\chi_{j} \sigma^{\mu} \bar{\lambda}^{\prime}_{k}) \phi_{l},
\qquad
f^{(9b)}_{jkl} (\chi_{j} \sigma^{\mu} \partial_{\mu}\bar{\lambda}^{\prime}_{k}) \phi_{l},
\nonumber \\
f^{(9c)}_{jkl} (\partial_{\mu}\lambda^{\prime}_{j} \sigma^{\mu} \bar{\chi}_{k}) \phi_{l},
\qquad
f^{(9d)}_{jkl} (\lambda^{\prime}_{j} \sigma^{\mu} \partial_{\mu}\bar{\chi}_{k}) \phi_{l},
\nonumber \\
f^{(9e)}_{jkl} (\partial_{\mu}\chi_{j} \sigma^{\mu} \bar{\lambda}^{\prime}_{k}) \phi^{\dagger}_{l},
\qquad
f^{(9f)}_{jkl} (\chi_{j} \sigma^{\mu} \partial_{\mu} \bar{\lambda}^{\prime}_{k}) \phi^{\dagger}_{l},
\nonumber \\
f^{(9g)}_{jkl} (\partial_{\mu}\lambda^{\prime}_{j} \sigma^{\mu} \bar{\chi}_{k}) \phi^{\dagger}_{l},
\qquad
f^{(9h)}_{jkl} (\lambda^{\prime}_{j} \sigma^{\mu} \partial_{\mu}\bar{\chi}_{k}) \phi^{\dagger}_{l}
\eea
and terms with one derivative on 
$\phi_{l}, \phi_{l}^{\dagger}$
which will be denoted generically by
$
\tilde{F}^{(9)};
$
\item
from the sector 
$
\phi \phi v_{\mu}
$
\be
f^{(10a)}_{jkl} \phi_{j} \partial_{\mu}\phi_{k} v_{l}^{\mu},
\qquad
f^{(10b)}_{jkl} \phi_{j} \partial_{\mu}\phi_{k}^{\dagger} v_{l}^{\mu},
\qquad
f^{(10c)}_{jkl} \phi_{j}^{\dagger} \partial_{\mu}\phi_{k}^{\dagger} v_{l}^{\mu},
\qquad
f^{(10d)}_{jkl} \phi_{j}^{\dagger} \partial_{\mu}\phi_{k} v_{l}^{\mu}
\ee
and terms with one derivative on 
$v^{\mu}_{l}$
which will be denoted generically by
$
\tilde{F}^{(10)};
$
\item
from the sector 
$
\phi \phi d
$
\be
f^{(11a)}_{jkl} \phi_{j} \phi_{k} d_{l},
\qquad
f^{(11b)}_{jkl} \phi_{j} \phi_{k}^{\dagger} d_{l},
\qquad
f^{(11c)}_{jkl} \phi_{j}^{\dagger} \phi_{k}^{\dagger} d_{l};
\ee
\item
from the sector
$
\phi\lambda\lambda:
$
\be
f^{(12a)}_{jkl} (\lambda_{j}^{\prime}\lambda^{\prime}_{k}) \phi_{l},
\qquad
f^{(12b)}_{jkl} (\lambda_{j}^{\prime}\lambda^{\prime}_{k}) \phi^{\dagger}_{l},
\qquad
f^{(12c)}_{jkl} (\bar{\lambda}_{j}^{\prime}\bar{\lambda}^{\prime}_{k}) \phi_{l},
\qquad
f^{(12d)}_{jkl} (\bar{\lambda}_{j}^{\prime}\bar{\lambda}^{\prime}_{k}) \phi^{\dagger}_{l};
\ee
\item
from the sector 
$
v_{\mu} v_{\nu} v_{\rho}
$
\be
f^{(13a)}_{jkl} v_{j}^{\mu} v_{k}^{\nu} \partial_{\nu}v_{l\mu} \qquad
f^{(13b)}_{jkl} v_{j}^{\mu} v_{k\mu} \partial_{\nu}v_{l}^{\nu}
\ee
\item
from the sector
$
v_{\mu}\lambda\lambda:
$
\be
f^{(14)}_{jkl} (\lambda_{j}^{\prime}\sigma_{\mu}\bar{\lambda}^{\prime}_{k}) v_{l}^{\mu}
\ee
\item
from the sector
$
\chi\lambda v_{\mu}:
$
\bea
f^{(15a)}_{jkl} (\partial^{\mu}\chi_{j} \sigma_{\mu\nu} \lambda^{\prime}_{k}) v^{\nu}_{l},
\qquad
f^{(15b)}_{jkl} (\chi_{j} \sigma_{\mu\nu} \partial^{\mu}\lambda^{\prime}_{k}) v^{\nu}_{l},
\nonumber \\
f^{(15c)}_{jkl} (\partial^{\mu}\bar{\chi}_{j} \bar{\sigma}_{\mu\nu} \bar{\lambda}^{\prime}_{k}) v^{\nu}_{l},
\qquad
f^{(15d)}_{jkl} (\bar{\chi}_{j} \bar{\sigma}_{\mu\nu} \partial^{\mu}\bar{\lambda}^{\prime}_{k}) v^{\nu}_{l},
\nonumber \\
f^{(15e)}_{jkl} (\partial_{\mu}\chi_{j} \lambda^{\prime}_{k}) v^{\mu}_{l},
\qquad
f^{(15f)}_{jkl} (\chi_{j} \partial_{\mu}\lambda^{\prime}_{k}) v^{\mu}_{l},
\nonumber \\
f^{(15g)}_{jkl} (\partial_{\mu}\bar{\chi}_{j} \bar{\lambda}^{\prime}_{k}) v^{\mu}_{l},
\qquad
f^{(15h)}_{jkl} (\bar{\chi}_{j} \partial_{\mu}\bar{\lambda}^{\prime}_{k}) v^{\mu}_{l}
\eea
and terms with one derivative on 
$v^{\mu}_{l}$
which will be denoted generically by
$
\tilde{F}^{(15)};
$
\item
from the sector
$
\chi\lambda d:
$
\be
f^{(16a)}_{jkl} (\chi_{j} \lambda^{\prime}_{k}) d_{l},
\qquad
f^{(16b)}_{jkl} (\bar{\chi}_{j} \bar{\lambda}^{\prime}_{k}) d_{l};
\ee
\item
from the sector
$
v_{\mu}v_{\nu} d:
$
\be
f^{(17)}_{jkl} v_{j\mu} v^{\mu}_{k} d_{l}.
\ee
\end{itemize}

II.) We proceed in the same way with the ghost terms. 
We have to consider terms of the type
$
A_{j} A_{k} A_{l}
$
where
$
A_{j} = C, \phi, \phi^{\dagger}, v_{\mu}, d, \chi, \lambda^{\prime}
$
and
$
A_{k} = u, v, g, \zeta
$
and
$
A_{l} = \tilde{u}, \tilde{v}, \tilde{g}, \tilde{\zeta}.
$
\begin{itemize}
\item 
from the sector 
$
Au\tilde{u}:
$
\be
g^{(1a)}_{jkl} v^{\mu}_{j} u_{k} \partial_{\mu}\tilde{u}_{l} \qquad
g^{(1b)}_{jkl} \partial_{\mu}v^{\mu}_{j} u_{k} \tilde{u}_{l} \qquad
g^{(1c)}_{jkl} d_{j} u_{k} \tilde{u}_{l} \quad
\ee
and terms with the derivative on
$
u_{k}
$
which will be generically denoted by
$
\tilde{G}^{(1)};
$
\item 
from the sector 
$
Av\tilde{v}:
$
\be
g^{(2a)}_{jkl} v^{\mu}_{j} v_{k} \partial_{\mu}\tilde{v}_{l} \qquad
g^{(2b)}_{jkl} \partial_{\mu}v^{\mu}_{j} v_{k} \tilde{v}_{l} \qquad
g^{(2c)}_{jkl} d_{j} v_{k} \tilde{v}_{l}
\ee
and terms with the derivative on
$
v_{k}
$
which will be generically denoted by
$
\tilde{G}^{(2)};
$
\item 
from the sector 
$
Av\tilde{u}:
$
\be
g^{(3a)}_{jkl} v^{\mu}_{j} v_{k} \partial_{\mu}\tilde{u}_{l} \qquad
g^{(3b)}_{jkl} \partial_{\mu}v^{\mu}_{j} v_{k} \tilde{u}_{l} \qquad
g^{(3c)}_{jkl} d_{j} v_{k} \tilde{u}_{l}
\ee
and terms with the derivative on
$
v_{k}
$
which will be generically denoted by
$
\tilde{G}^{(3)};
$
\item 
from the sector 
$
Au\tilde{v}:
$
\be
g^{(4a)}_{jkl} v^{\mu}_{j} \partial_{\mu}u_{k} \tilde{v}_{l} \qquad
g^{(4b)}_{jkl} \partial_{\mu}v^{\mu}_{j} u_{k} \tilde{v}_{l} \qquad
g^{(4c)}_{jkl} d_{j} u_{k} \tilde{v}_{l}
\ee
and terms with the derivative on
$
\tilde{v}_{l}
$
which will be generically denoted by
$
\tilde{G}^{(4)};
$
\item 
from the sector 
$
Ag\tilde{u}:
$
\be
g^{(5a)}_{jkl} \phi_{j}g_{k}\tilde{u}_{l} \qquad
g^{(5b)}_{jkl} \phi^{\dagger}_{j}g_{k}\tilde{u}_{l} \qquad
g^{(5c)}_{jkl} \phi_{j}g^{\dagger}_{k}\tilde{u}_{l} \qquad
g^{(5d)}_{jkl} \phi^{\dagger}_{j}g^{\dagger}_{k}\tilde{u}_{l}
\ee
\item 
from the sector 
$
Ag\tilde{v}:
$
\be
g^{(6a)}_{jkl} \phi_{j} g_{k} \tilde{v}_{l} \qquad
g^{(6b)}_{jkl} \phi^{\dagger}_{j} g_{k} \tilde{v}_{l} \qquad
g^{(6c)}_{jkl} \phi_{j} g^{\dagger}_{k} \tilde{v}_{l} \qquad
g^{(6d)}_{jkl} \phi^{\dagger}_{j} g^{\dagger}_{k} \tilde{v}_{l}
\ee
\item 
from the sector 
$
Au\tilde{g}:
$
\be
g^{(7a)}_{jkl} \phi_{j} u_{k} \tilde{g}_{l} \qquad
g^{(7b)}_{jkl} \phi^{\dagger}_{j} u_{k} \tilde{g}_{l} \qquad
g^{(7c)}_{jkl} \phi_{j} u_{k} \tilde{g}^{\dagger}_{l} \qquad
g^{(7d)}_{jkl} \phi^{\dagger}_{j} u_{k} \tilde{g}^{\dagger}_{l}
\ee
\item 
from the sector 
$
Av\tilde{g}:
$
\be
g^{(8a)}_{jkl} \phi_{j} v_{k} \tilde{g}_{l} \qquad
g^{(8b)}_{jkl} \phi^{\dagger}_{j} v_{k} \tilde{g}_{l} \qquad
g^{(8c)}_{jkl} \phi_{j} v_{k} \tilde{g}^{\dagger}_{l} \qquad
g^{(8d)}_{jkl} \phi^{\dagger}_{j} v_{k} \tilde{g}^{\dagger}_{l}
\ee
\item 
from the sector
$
Ag\tilde{g}:
$
\be
g^{(9a)}_{jkl} C_{j} g_{k} \tilde{g}_{l} \qquad
g^{(9b)}_{jkl} C_{j} g_{k}^{\dagger} \tilde{g}_{l} \qquad
g^{(9c)}_{jkl} C_{j} g_{k} \tilde{g}_{l}^{\dagger} \qquad
g^{(9d)}_{jkl} C_{j} g^{\dagger}_{k} \tilde{g}^{\dagger}_{l}
\ee
\item 
from the sector
$
A\zeta\tilde{u}:
$
\bea
g^{(10a)}_{jkl} (\lambda^{\prime}_{j}\zeta_{k}) \tilde{u}_{l} \qquad
g^{(10b)}_{jkl} (\bar{\lambda}^{\prime}_{j}\bar{\zeta}_{k}) \tilde{u}_{l} \qquad
\nonumber \\
g^{(10c)}_{jkl} (\partial_{\mu}\chi_{j}\sigma^{\mu}\bar{\zeta}_{k}) \tilde{u}_{l} \qquad
g^{(10d)}_{jkl} (\chi_{j}\sigma^{\mu}\partial_{\mu}\bar{\zeta}_{k}) \tilde{u}_{l} \qquad
\nonumber \\
g^{(10e)}_{jkl} (\partial_{\mu}\zeta_{k}\sigma^{\mu}\bar{\chi}_{j}) \tilde{u}_{l} \qquad
g^{(10f)}_{jkl} (\zeta_{k}\sigma^{\mu}\partial_{\mu}\bar{\chi}_{j}) \tilde{u}_{l} \qquad
\eea
and terms with the derivative on
$
\tilde{u}_{l}
$
which will be generically denoted by
$
\tilde{G}^{(10)};
$
\item 
from the sector
$
A\zeta\tilde{v}:
$
\bea
g^{(11a)}_{jkl} (\lambda^{\prime}_{j}\zeta_{k}) \tilde{v}_{l} \qquad
g^{(11b)}_{jkl} (\bar{\lambda}^{\prime}_{j}\bar{\zeta}_{k}) \tilde{v}_{l} \qquad
\nonumber \\
g^{(11c)}_{jkl} (\partial_{\mu}\chi_{j}\sigma^{\mu}\bar{\zeta}_{k}) \tilde{v}_{l} \qquad
g^{(11d)}_{jkl} (\chi_{j}\sigma^{\mu}\partial_{\mu}\bar{\zeta}_{k}) \tilde{v}_{l} \qquad
\nonumber \\
g^{(11e)}_{jkl} (\partial_{\mu}\zeta_{k}\sigma^{\mu}\bar{\chi}_{j}) \tilde{v}_{l} \qquad
g^{(11f)}_{jkl} (\zeta_{k}\sigma^{\mu}\partial_{\mu}\bar{\chi}_{j}) \tilde{v}_{l}
\eea
and terms with the derivative on
$
\tilde{v}_{l}
$
which will be generically denoted by
$
\tilde{G}^{(11)};
$
\item 
from the sector
$
Au\tilde{\zeta}:
$
\bea
g^{(12a)}_{jkl} (\lambda^{\prime}_{j}\tilde{\zeta}_{l}) u_{k} \qquad
g^{(12b)}_{jkl} (\bar{\lambda}^{\prime}_{j}\bar{\tilde{\zeta}}_{l}) u_{k} \qquad
\nonumber \\
g^{(12c)}_{jkl} (\partial_{\mu}\chi_{j}\sigma^{\mu}\bar{\tilde{\zeta}}_{l}) u_{k} \qquad
g^{(12d)}_{jkl} (\chi_{j}\sigma^{\mu}\partial_{\mu}\bar{\tilde{\zeta}}_{l}) u_{k} \qquad
\nonumber \\
g^{(12e)}_{jkl} (\partial_{\mu}\tilde{\zeta}_{l}\sigma^{\mu}\bar{\chi}_{j}) u_{k} \qquad
g^{(12f)}_{jkl} (\tilde{\zeta}_{l}\sigma^{\mu}\partial_{\mu}\bar{\chi}_{j}) u_{k}
\eea
and terms with the derivative on
$
u_{k}
$
which will be generically denoted by
$
\tilde{G}^{(12)};
$
\item 
from the sector
$
Av\tilde{\zeta}:
$
\bea
g^{(13a)}_{jkl} (\lambda^{\prime}_{j}\tilde{\zeta}_{k}) v_{l} \qquad
g^{(13b)}_{jkl} (\bar{\lambda}^{\prime}_{j}\bar{\tilde{\zeta}}_{l}) v_{l} \qquad
\nonumber \\
g^{(13c)}_{jkl} (\partial_{\mu}\chi_{j}\sigma^{\mu}\bar{\tilde{\zeta}}_{l}) v_{k} \qquad
g^{(13d)}_{jkl} (\chi_{j}\sigma^{\mu}\partial_{\mu}\bar{\tilde{\zeta}}_{l}) v_{k} \qquad
\nonumber \\
g^{(13e)}_{jkl} (\partial_{\mu}\tilde{\zeta}_{l}\sigma^{\mu}\bar{\chi}_{j}) v_{k} \qquad
g^{(13f)}_{jkl} (\tilde{\zeta}_{l}\sigma^{\mu}\partial_{\mu}\bar{\chi}_{j}) v_{k}
\eea
and terms with the derivative on
$
v_{k}
$
which will be generically denoted by
$
\tilde{G}^{(13)};
$
\item 
from the sector
$
A\zeta\tilde{\zeta}:
$
\bea
g^{(14a)}_{jkl} \phi_{j} (\zeta_{k}\tilde{\zeta}_{l}) \qquad
g^{(14b)}_{jkl} \phi^{\dagger}_{j} (\zeta_{k}\tilde{\zeta}_{l}) \qquad
g^{(14c)}_{jkl} \phi_{j} (\bar{\zeta}_{k}\bar{\tilde{\zeta}}_{l}) \qquad
g^{(14d)}_{jkl} \phi^{\dagger}_{j} (\bar{\zeta}_{k}\bar{\tilde{\zeta}}_{l})
\nonumber \\
g^{(14e)}_{jkl} C_{j} (\partial_{\mu}\zeta_{k}\sigma^{\mu}\bar{\tilde{\zeta}}_{l}) \qquad
g^{(14f)}_{jkl} C_{j} (\zeta_{k}\sigma^{\mu}\partial_{\mu}\bar{\tilde{\zeta}}_{l})
\nonumber \\
g^{(14g)}_{jkl} C_{j} (\partial_{\mu}\tilde{\zeta}_{l}\sigma^{\mu}\bar{\zeta}_{k}) \qquad
g^{(14h)}_{jkl} C_{j} (\tilde{\zeta}_{l}\sigma^{\mu}\partial_{\mu}\bar{\zeta}_{k})
\nonumber \\
g^{(14i)}_{jkl} v_{j}^{\mu}~(\zeta_{k}\sigma_{\mu}\bar{\tilde{\zeta}}_{l}) \qquad
g^{(14j)}_{jkl} v_{j}^{\mu}~(\tilde{\zeta}_{l}\sigma_{\mu}\bar{\zeta}_{k})~~ 
\eea
and terms with the derivative on
$
C_{j}
$
which will be generically denoted by
$
\tilde{G}^{(14)};
$

\item
from the sector
$
A\zeta\tilde{g}:
$
\be
g^{(15a)}_{jkl} (\chi_{j}\zeta_{k}) \tilde{g}_{l} \qquad
g^{(15b)}_{jkl} (\chi_{j}\zeta_{k}) \tilde{g}_{l}^{\dagger} \qquad
g^{(15c)}_{jkl} (\bar{\chi}_{j}\bar{\zeta}_{k}) \tilde{g}_{l} \qquad
g^{(15d)}_{jkl} (\bar{\chi}_{j}\bar{\zeta}_{k}) \tilde{g}_{l}^{\dagger}
\ee
\item
from the sector
$
Ag\tilde{\zeta}:
$
\be
g^{(16a)}_{jkl} (\chi_{j}\tilde{\zeta}_{l}) g_{k} \qquad
g^{(16b)}_{jkl} (\chi_{j}\tilde{\zeta}_{l}) g_{k}^{\dagger} \qquad
g^{(16c)}_{jkl} (\bar{\chi}_{j}\bar{\tilde{\zeta}}_{l}) g_{k} \qquad
g^{(16d)}_{jkl} (\bar{\chi}_{j}\bar{\tilde{\zeta}}_{l}) g_{k}^{\dagger}
\ee
\end{itemize}

Some of the possible terms have been discarded according to the ``magic"
formula. The coefficients are subject to various obvious (anti)symmetry
properties. 

III.) We have total divergence terms 
$
t^{(j)}_{\mu}~,j = 1,\dots,20 
$
in the sectors
$$
f^{(1)} - f^{(4)}, f^{(6)} -f^{(10)},f^{(13)}, f^{(15)}, g^{(1)} - g^{(4)}, g^{(10)} - g^{(14)}
$$
respectively.

IV.) We also have the co-boundary terms
$
d_{Q}b
$
to eliminate some of the previous expressions; we list the possible expressions
$b$:
\begin{itemize}
\item
of the type
$
A A^{\prime} \tilde{u}:
$
\bea
b^{(1a)}_{jkl}~\phi_{j} \phi_{k} \tilde{u}_{l} \qquad 
b^{(1b)}_{jkl}~\phi^{\dagger}_{j} \phi^{\dagger}_{k} \tilde{u}_{l} \qquad 
b^{(1c)}_{jkl}~\phi_{j} \phi^{\dagger}_{k} \tilde{u}_{l} \qquad 
b^{(1d)}_{jkl}~v^{\mu}_{j} v_{k\mu} \tilde{u}_{l} \qquad
b^{(1e)}_{jkl}~C_{j} d_{k} \tilde{u}_{l} 
\nonumber \\
b^{(1f)}_{jkl}~(\chi_{j}\lambda^{\prime}_{k}) \tilde{u}_{l} \qquad 
b^{(1g)}_{jkl}~(\bar{\chi}_{j}\bar{\lambda}^{\prime}_{k}) \tilde{u}_{l} \qquad 
b^{(1h)}_{jkl}~(\partial_{\mu}\chi_{j}\sigma^{\mu}\bar{\chi}_{k}) \tilde{u}_{l} \qquad 
b^{(1i)}_{jkl}~(\chi_{j}\sigma^{\mu}\partial_{\mu}\bar{\chi}_{k}) \tilde{u}_{l}
\nonumber \\
b^{(1j)}_{jkl}~C_{j} \partial_{\mu}v^{\mu}_{k} \tilde{u}_{l} \qquad
b^{(1k)}_{jkl}~C_{j} v^{\mu}_{k} \partial_{\mu}\tilde{u}_{l}
\eea
\item
of the type
$
A A^{\prime} \tilde{v}:
$
\bea
b^{(2a)}_{jkl}~\phi_{j} \phi_{k} \tilde{v}_{l} \qquad 
b^{(2b)}_{jkl}~\phi^{\dagger}_{j} \phi^{\dagger}_{k} \tilde{v}_{l} \qquad 
b^{(2c)}_{jkl}~\phi_{j} \phi^{\dagger}_{k} \tilde{v}_{l} \qquad 
b^{(2d)}_{jkl}~v^{\mu}_{j} v_{k\mu} \tilde{v}_{l} \qquad 
b^{(2e)}_{jkl}~C_{j} d_{k} \tilde{v}_{l}
\nonumber \\
b^{(2f)}_{jkl}~(\chi_{j}\lambda^{\prime}_{k}) \tilde{v}_{l} \qquad 
b^{(2g)}_{jkl}~(\bar{\chi}_{j}\bar{\lambda}^{\prime}_{k}) \tilde{v}_{l} \qquad 
b^{(2h)}_{jkl}~(\partial_{\mu}\chi_{j}\sigma^{\mu}\bar{\chi}_{k}) \tilde{v}_{l} \qquad 
b^{(2i)}_{jkl}~(\chi_{j}\sigma^{\mu}\partial_{\mu}\bar{\chi}_{k}) \tilde{v}_{l}
\nonumber \\
b^{(2j)}_{jkl}~C_{j} \partial_{\mu}v^{\mu}_{k} \tilde{v}_{l} \qquad
b^{(2k)}_{jkl}~C_{j} v^{\mu}_{k} \partial_{\mu}\tilde{v}_{l}
\eea
\item
of the type
$
A A^{\prime} \tilde{g}:
$
\bea
b^{(3a)}_{jkl}~C_{j} \phi_{k} \tilde{g}_{l} \qquad 
b^{(3b)}_{jkl}~C_{j} \phi^{\dagger}_{k} \tilde{g}_{l} \qquad 
b^{(3c)}_{jkl}~C_{j} \phi_{k} \tilde{g}^{\dagger}_{l} \qquad 
b^{(3d)}_{jkl}~C_{j} \phi^{\dagger}_{k} \tilde{g}^{\dagger}_{l}
\nonumber \\
b^{(3e)}_{jkl}~(\chi_{j}\chi_{k}) \tilde{g}_{l} \qquad 
b^{(3f)}_{jkl}~(\chi_{j}\chi_{k}) \tilde{g}^{\dagger}_{l} \qquad 
b^{(3g)}_{jkl}~(\bar{\chi}_{j}\bar{\chi}_{k}) \tilde{g}_{l} \qquad 
b^{(3h)}_{jkl}~(\bar{\chi}_{j}\bar{\chi}_{k}) \tilde{g}^{\dagger}_{l}
\eea
\item
of the type
$
A A^{\prime} \tilde{\zeta}:
$
\bea
b^{(4a)}_{jkl}~\phi_{j} (\chi_{k} \tilde{\zeta}_{l}) \qquad 
b^{(4b)}_{jkl}~\phi^{\dagger}_{j} (\chi_{k} \tilde{\zeta}_{l}) \qquad 
b^{(4c)}_{jkl}~\phi_{j} (\bar{\chi}_{k} \bar{\tilde{\zeta}}_{l}) \qquad 
b^{(4d)}_{jkl}~\phi^{\dagger}_{j} \bar{\chi}_{k} \bar{\tilde{\zeta}}_{l}) \qquad 
\nonumber \\
b^{(4e)}_{jkl}~C_{j} (\partial_{\mu}\chi_{k}\sigma^{\mu}\bar{\tilde{\zeta}}_{l}) \qquad 
b^{(4f)}_{jkl}~C_{j} (\chi_{k} \sigma^{\mu} \partial_{\mu}\bar{\tilde{\zeta}}_{l}) \qquad
\nonumber \\
b^{(4g)}_{jkl}~C_{j} (\partial_{\mu}\tilde{\chi}_{l}\sigma^{\mu}\bar{\zeta}_{k}) \qquad 
b^{(4h)}_{jkl}~C_{j} (\tilde{\chi}_{k} \sigma^{\mu} \partial_{\mu}\bar{\zeta}_{l})
\nonumber \\
b^{(4i)}_{jkl}~C_{j} (\lambda^{\prime}_{k} \tilde{\zeta}_{l}) \qquad 
b^{(4j)}_{jkl}~C_{j} (\bar{\lambda}^{\prime}_{k} \bar{\tilde{\zeta}}_{l}) \qquad 
b^{(4k)}_{jkl}~v_{j}^{\mu} (\chi_{k} \sigma_{\mu} \bar{\tilde{\zeta}}_{l}) \qquad
b^{(4l)}_{jkl}~v_{j}^{\mu} (\tilde{\zeta}_{k} \sigma_{\mu} \bar{\chi}_{l})
\eea
\item
tri-linear in the ghost fields:
\be
b^{(5a)}_{jkl}~u_{j}\tilde{u}_{k}\tilde{u}_{l} \quad
b^{(5b)}_{jkl}~u_{j}\tilde{u}_{k}\tilde{v}_{l} \quad
b^{(5c)}_{jkl}~u_{j}\tilde{v}_{k}\tilde{v}_{l} \quad
b^{(5d)}_{jkl}~v_{j}\tilde{u}_{k}\tilde{u}_{l} \quad
b^{(5e)}_{jkl}~v_{j}\tilde{u}_{k}\tilde{v}_{l} \quad
b^{(5f)}_{jkl}~v_{j}\tilde{v}_{k}\tilde{v}_{l}.
\ee
\end{itemize}

V.) Now we eliminate terms of the type $f$ and $g$ using total divergences. We have
to pay care to the order in which we proceed because if we use a total 
divergence (or a co-boundary) we must not modify terms which have already been 
fixed. We proceed as follows: first we use 
\be
t^{(10)}_{\mu} \equiv t_{jkl} v_{j}^{\nu} v_{k\nu} v_{l\mu} \qquad 
t_{jkl} = t_{kjl};
\ee
because
\be
\partial^{\mu}t^{(10)}_{\mu} = t_{jkl} ( 2\partial^{\mu}v_{j}^{\nu} v_{k\nu} v_{l\mu}
+ v_{j}^{\nu} v_{k\nu} \partial_{\mu}v_{l}^{\mu} )
\ee
it is possible to take
\be
f^{(13a)}_{jkl} = - f^{(13a)}_{lkj};
\label{a1}
\ee
For simplicity we denote from now on:
$
f^{(13)}_{jkl} \equiv f^{(13a)}_{jkl}.
$

We find convenient to describe the various redefinitions in the following table:
\vskip 0.5cm
\begin{tabular}{|c|c|c|c|}
\hline
\hline
Nr. crt. & $t_{\mu}, b$  &  Restrictions & Modified Terms \\  \hline \hline
1  & $t^{(10)}$ &  $f^{(13a)}_{jkl} = - (j \longleftrightarrow l)$  &  $f^{(13b)}_{jkl}$ \\  \hline
2  & $b^{(1d)}$ & $f^{(13b)} = 0$  & $\tilde{G}^{(1)}$      \\ \hline
3  & $t^{(12)}$ & $\tilde{G}^{(1)} = 0$  &  $g^{(1a)},g^{(1b)} $  \\ \hline
4  & $b^{(1a)}$ & $g^{(5c)}_{jkl} = - (j \longleftrightarrow k)$  & $\tilde{F}^{(10)}$  \\ \hline
5  & $b^{(1b)}$ & $g^{(5b)}_{jkl} = - (j \longleftrightarrow k)$   & $\tilde{F}^{(10)}$  \\ \hline
6  & $b^{(1c)}$ & $g^{(5d)} = 0$   & $\tilde{F}^{(10)}, g^{(5a)}$   \\ \hline
7  & $b^{(1e)}$ & $g^{(3c)} = 0$   & $f^{(3b)}$               \\ \hline
8  & $b^{(1f)}$ & $g^{(10a)} = 0$  & $\tilde{F}^{(15)}$             \\  \hline
9  & $b^{(1g)}$ & $g^{(10b)} = 0$  & $\tilde{F}^{(15)}$              \\  \hline
10 & $t^{(16)}_{\mu}$ & $\tilde{G}^{(10)} = 0$  & $g^{(10c)},\dots, g^{(10f)}$ \\ \hline
11 & $b^{(1h)}$ & $g^{(10c)} = 0$  & $g^{(10e)}, \tilde{F}^{(7)}$   \\ \hline
12 & $b^{(1i)}$ & $g^{(10f)} = 0$  & $g^{(10d)}, \tilde{F}^{(7)}$  \\ \hline
13 & $t^{(2)}_{\mu}$ & $\tilde{F}^{(2)}$  & $f^{(2a)},\dots, f^{(2d)}$  \\ \hline
14 & $t^{(14)}_{\mu}$& $\tilde{G}^{(3)}$  & $g^{(3a)}, g^{(3b)}$  \\  \hline
15 & $b^{(1j)}$ & $f^{(2d)} = 0$  & $g^{(3b)}$  \\ \hline
16 & $b^{(1h)}$ & $f^{(2b)} = 0$  & $g^{(3a)}$, {\rm total div} \\ \hline
17 & $b^{(2a)}$ & $g^{(6c)}_{jkl} = - (j \longleftrightarrow k)$  & $f^{(11a)}$  \\ \hline
18 & $b^{(2b)}$ & $g^{(6b)}_{jkl} = - (j \longleftrightarrow k)$  & $f^{(11c)}$  \\ \hline
19 & $b^{(2c)}$ & $g^{(6d)} = 0$ & $g^{(6a)}, f^{(11b)}$  \\ \hline
20 & $t^{(15)}_{\mu}$ & $\tilde{G}^{(4)} = 0$ & $g^{(4a)}, g^{(4b)}$  \\ \hline
\end{tabular}

\begin{tabular}{|c|c|c|c|}
\hline
\hline
Nr. crt. & $t_{\mu}, b$  &  Restrictions & Modified Terms \\  \hline \hline
21 & $b^{(2d)}$ & $g^{(4a)}_{jkl} =  - (j \leftrightarrow k)$  & $f^{(17)}$  \\ \hline
22 & $b^{(2e)}$ & $f^{(5)} = 0$  &  $g^{(2c)}$ \\ \hline
23 & $b^{(2f)}$ & $g^{(11a)} = 0$  & $f^{(16a)}$  \\ \hline
24 & $b^{(2g)}$ & $g^{(11b)} = 0$  & $f^{(16b)}$  \\ \hline
25 & $t^{(17)}_{\mu}$  & $\tilde{G}^{(11)} = 0$  & $g^{(11c)} - g^{(11f)}$  \\ \hline
26 & $b^{(2h)}$ & $g^{(11c)} = 0$  & $g^{(11e)}, f^{(8a)}$  \\ \hline
27 & $b^{(2i)}$ & $g^{(11f)} = 0$  & $g^{(11d)}, f^{(8b)}$  \\ \hline
28 & $t^{(13)}_{\mu}$  & $\tilde{G}^{(2)} = 0$  & $g^{(2a)}, g^{(2b)}$  \\ \hline
29 & $t^{(3)}_{\mu}$   & $\tilde{F}^{(3)} = 0$  & $f^{(3a)}, f^{(3b)}$  \\ \hline
30 & $b^{(2j)}$ & $f^{(3b)} = 0$  & $g^{(2b)}$ \\ \hline
31 & $b^{(2h)}$ & $f^{(3a)} = 0$  & $g^{(2a)}$, {\rm tot div} \\ \hline
32 & $b^{(3a)} - b^{(3d)}$ & $g^{(8a)},\dots, g^{(8d)} = 0$ & $g^{(9a)} - g^{(9d)}$ \\ \hline
33 & $b^{(3e)} - b^{(3h)}$ & $g^{(15a,b,c,d)}_{jkl} = - (j \leftrightarrow k)$ &  \\ \hline
34 & $b^{(4a)} - b^{(4d)}$ & $g^{(14a)} - g^{(14d)} = 0$ & $g^{(16a)} - g^{(16d)}, f^{(9b)}, f^{(9f)}, f^{(9c)}, f^{(9g)}$  \\ \hline
35 & $t^{(20)}_{\mu}$ & $\tilde{G}^{(14)} = 0$ & $g^{(14e)} - g^{(14h)}$ \\ \hline
36 & $b^{(4e)} - b^{(4h)}$ & $g^{(14e)} - g^{(14h)} = 0$ & $g^{(13c)} - g^{(13f)}, f^{(1a)}, f^{(1b)}$, {\rm total div} \\ \hline
37 & $b^{(4k)}, b^{(4l)}$ & $g^{(14i)} = g^{(14j)} = 0$ &  $\tilde{G}^{(12)}, f^{(15f)}, f^{(15b)}, f^{(15h)}, f^{(15d)}$ \\ \hline
38 & $t^{(18)}_{\mu}$ & $\tilde{G}^{(12)} = 0$ &  $g^{(12c)} - g^{(12f)}$ \\ \hline
39 & $t^{(19)}_{\mu}$ & $\tilde{G}^{(13)} = 0$ &  $g^{(13c)} - g^{(13f)}$ \\ \hline
40 & $b^{(4i)}$  & $g^{(13a)} = 0$ &  $f^{(4a)}$ \\ \hline
41 & $b^{(4j)}$  & $g^{(13b)} = 0$ &  $f^{(4b)}$ \\ \hline
42 & $b^{(5a)}$  & $g^{(1b)}_{jkl} = (j \leftrightarrow l)$ &  \\ \hline
43 & $b^{(5b)}$  & $g^{(1c)} = 0$ &  $g^{(4b)}$\\ \hline
44 & $b^{(5c)}$  & $g^{(4c)}_{jkl} = (j \leftrightarrow l)$ &  \\ \hline
45 & $b^{(5d)}$  & $g^{(3b)}_{jkl} = (j \leftrightarrow l)$ &  \\ \hline
46 & $b^{(5e})$  & $g^{(3c)} = 0$ & $g^{(2b)}$ \\ \hline
47 & $b^{(5f)}$  & $g^{(2c)}_{jkl} = (j \leftrightarrow l)$ &  \\ \hline
48 & $t^{(1)}_{\mu}, t^{(4)}_{\mu} $ & $\tilde{F}^{(1)} = \tilde{F}^{(4)} = 0$ & 
$f^{(1a)}, f^{(1b)}, f^{(4a)}, f^{(4b)}$ \\ \hline
49 & $t^{(5)}_{\mu}, t^{(6)}_{\mu} $ & $\tilde{F}^{(6)} = \tilde{F}^{(7)} = 0$ & 
$f^{(6a)} - f^{(6d)}, f^{(7a)} - f^{(7d)}$ \\ \hline
50 & $t^{(7)}_{\mu}, t^{(8)}_{\mu} $ & $\tilde{F}^{(8)} = \tilde{F}^{(9)} = 0$ & 
$f^{(8a)}, f^{(8b)}, f^{(9a)} - f^{(9h)}$ \\ \hline
51 & $t^{(9)}_{\mu}, t^{(11)}_{\mu} $ & $\tilde{F}^{(10)} = \tilde{F}^{(15)} = 0$ & 
$f^{(10a)} - f^{(10d)}, f^{(15a)} - f^{(15h)}$ \\ \hline
\end{tabular}
\vskip 0.5cm
Using (\ref{gauge4}) we can compute the expression
$
d_{Q}T.
$
We exhibit it in the form 
\be
d_{Q}t = d_{Q}t^{(3)} + i u_{j} A_{j} + i v_{j} B_{j} 
+ g_{j} A^{\prime}_{j} +  g^{\dagger}_{j} B^{\prime}_{j}
+ i \zeta_{j} X_{j} + i \bar{\zeta}_{j} \bar{X}^{\prime}_{j} + {\rm total~div}
\ee
where the first term is tri-linear in the ghost fields and the expressions
$
A_{j},A^{\prime}_{j},B_{j},B^{\prime}_{j},X_{j},X^{\prime}_{j}
$
are independent of the ghost fields. We impose (\ref{gauge}) and note that the first term above must be a total divergence by himself. The explicit expression is
\bea
d_{Q}t^{(3)} = - g^{(5a)}_{jkl} g^{\dagger}_{j} g_{k} \tilde{u}_{l}
+ g^{(5b)}_{jkl} g_{j} g_{k} \tilde{u}_{l}
- g^{(5c)}_{jkl} g^{\dagger}_{j} g^{\dagger}_{k}\tilde{u}_{l} 
\nonumber \\
- g^{(6a)}_{jkl} g^{\dagger}_{j} g_{k} \tilde{v}_{l}
+ g^{(6b)}_{jkl} g_{j} g_{k} \tilde{v}_{l}
- g^{(6c)}_{jkl} g^{\dagger}_{j} g^{\dagger}_{k} \tilde{v}_{l}
\nonumber \\
- g^{(7a)}_{jkl} g^{\dagger}_{j} u_{k} \tilde{g}_{l}
+ g^{(7b)}_{jkl} g_{j} u_{k} \tilde{g}_{l}
- g^{(7c)}_{jkl} g^{\dagger}_{j} u_{k} \tilde{g}^{\dagger}_{l}
+ g^{(7d)}_{jkl} g_{j} u_{k} \tilde{g}^{\dagger}_{l}
\nonumber \\
+ i~g^{(9a)}_{jkl} v_{j} g_{k} \tilde{g}_{l}
+ i~g^{(9b)}_{jkl} v_{j} g^{\dagger}_{k} \tilde{g}_{l}
+ i~g^{(9c)}_{jkl} v_{j} g_{k} \tilde{g}^{\dagger}_{l}
+ i~g^{(9d)}_{jkl} v_{j} g^{\dagger}_{k} \tilde{g}^{\dagger}_{l}
\nonumber \\
+ 2 i~g^{(10d)}_{jkl} (\zeta_{j}\sigma^{\mu}\partial_{\mu}\bar{\zeta}_{k}) \tilde{u}_{l} 
- 2 i~g^{(10e)}_{jkl} (\partial_{\mu}\zeta_{j}\sigma^{\mu}\bar{\zeta}_{k}) \tilde{u}_{l} 
\nonumber \\
+ 2 i~g^{(11d)}_{jkl} (\zeta_{j}\sigma^{\mu}\partial_{\mu}\bar{\zeta}_{k}) \tilde{v}_{l} 
- 2 i~g^{(11e)}_{jkl} (\partial_{\mu}\zeta_{j}\sigma^{\mu}\bar{\zeta}_{k}) \tilde{v}_{l} 
\nonumber \\
+ 2 i~g^{(12c)}_{jkl}~
(\partial_{\mu}\zeta_{j}\sigma^{\mu}\bar{\tilde{\zeta}}_{k}) u_{k}
+ 2 i~g^{(12d)}_{jkl}~
(\zeta_{j}\sigma^{\mu}\partial_{\mu}\bar{\tilde{\zeta}}_{k}) u_{k}
\nonumber \\
- 2i~g^{(12e)}_{jkl} (\partial_{\mu}\tilde{\zeta}_{l}\sigma^{\mu}\bar{\zeta}_{j}) u_{k}
- 2 i~g^{(12f)}_{jkl} (\tilde{\zeta}_{j}\sigma^{\mu}\partial_{\mu}\bar{\zeta}_{k}) u_{k}
\nonumber \\
+ 2 i~g^{(13c)}_{jkl} (\partial_{\mu}\zeta_{j}\sigma^{\mu}\bar{\tilde{\zeta}}_{l}) v_{k}
+ 2 i~g^{(13d)}_{jkl} (\zeta_{j}\sigma^{\mu}\partial_{\mu}\bar{\tilde{\zeta}}_{l}) v_{k}
\nonumber \\
- 2i~g^{(13e)}_{jkl} (\partial_{\mu}\tilde{\zeta}_{l}\sigma^{\mu}\bar{\zeta}_{j}) v_{k}
- 2 i~g^{(13f)}_{jkl} (\tilde{\zeta}_{l}\sigma^{\mu}\partial_{\mu}\bar{\zeta}_{j}) v_{k}
\nonumber \\
+ 2 i~g^{(15a)}_{jkl} (\zeta_{j}\zeta_{k}) \tilde{g}_{l}
+ 2 i~g^{(15b)}_{jkl} (\zeta_{j}\zeta_{k}) \tilde{g}_{l}^{\dagger}
- 2 i~g^{(15c)}_{jkl} (\bar{\zeta}_{j}\bar{\zeta}_{k}) \tilde{g}_{l}
- 2 i~g^{(15d)}_{jkl} (\bar{\zeta}_{j}\bar{\zeta}_{k}) \tilde{g}_{l}^{\dagger}
\nonumber \\
+ 2 i~g^{(16a)}_{jkl} (\zeta_{j}\tilde{\zeta}_{l}) g_{k}
+ 2 i~g^{(16b)}_{jkl} (\zeta_{j}\tilde{\zeta}_{l}) g_{k}^{\dagger}
- 2 i~g^{(16c)}_{jkl} (\bar{\zeta}_{j}\bar{\tilde{\zeta}}_{l}) g_{k}
- 2 i~g^{(16d)}_{jkl} (\bar{\zeta}_{j}\bar{\tilde{\zeta}}_{l}) g_{k}^{\dagger}
\eea
and it is easy to see that
$
d_{Q}t^{(3)}
$
is a total divergence {\it iff} it is identically zero. This amounts to
\bea
g^{(p)} = 0 \qquad p = 5,6,7,9,10,11,15,16
\nonumber \\
g^{(12c)}_{jkl} = \quad\dots\quad = g^{(12f)}_{jkl} = 0
\nonumber \\
g^{(13c)}_{jkl} = 0 \quad \cdots \quad = g^{(13f)}_{jkl} = 0.
\eea

The expressions
$
A_{j},A^{\prime}_{j},B_{j},B^{\prime}_{j},X_{j},X^{\prime}_{j}
$
have the following form 
\bea
A_{j} = - 2 f^{(13)}_{jkl}~\partial^{\nu}v^{\mu}_{k}~\partial_{\mu}v_{l\nu}
\nonumber \\
+ (- f^{(13)}_{jkl} + f^{(13)}_{lkj} + f^{(13)}_{klj} + g^{(1a)}_{kjl})~
v^{\mu}_{k}~\partial_{\mu}\partial_{\nu}v^{\nu}_{l}
\nonumber \\
+ (f^{(13)}_{lkj} + g^{(1b)}_{kjl})~
\partial_{\mu}v^{\mu}_{k}~\partial_{\nu}v_{l}^{\nu}
\nonumber \\
- \left(f^{(14)}_{lkj} - {i\over 2}~g^{(12b)}_{kjl}\right)~
(\partial_{\mu}\lambda^{\prime}_{l}~\sigma^{\mu}~\bar{\lambda}_{k}^{\prime})
- \left(f^{(14)}_{lkj} + {i\over 2}~g^{(12a)}_{ljk}\right)~
(\lambda^{\prime}_{l}~\sigma^{\mu}~\partial_{\mu}\bar{\lambda}_{k}^{\prime})
\nonumber \\
- (f^{(15a)}_{klj} - f^{(15b)}_{klj})~
(\partial_{\mu}\chi_{l}~\sigma^{\mu\nu}~\partial_{\nu}\lambda_{k}^{\prime})
\nonumber \\
- (f^{(15c)}_{lkj} - f^{(15d)}_{lkj})~(\partial_{\mu}\bar{\chi}_{l}~\sigma^{\mu\nu}~
\partial_{\nu}\bar{\lambda}_{k}^{\prime})
\nonumber \\
- 2~(f^{(17)}_{jkl} - g^{(4a)}_{klj} + g^{(4b)}_{klj})~v^{\mu}_{k}~\partial_{\mu}d_{l}
- 2~(f^{(17)}_{jkl} - g^{(4a)}_{klj})~\partial_{\mu}v^{\mu}_{k}~d_{l}
- 2~g^{(4c)}_{klj}~d_{k}~d_{l}
\eea
\bea
B_{j} = f^{(1a)}_{jkl}~
(\partial_{\mu}\chi_{k}~\sigma^{\mu\nu}~\partial_{\nu}\lambda^{\prime}_{l})
+ f^{(1b)}_{jkl}~
(\partial_{\mu}\bar{\chi}_{k}~\bar{\sigma}^{\mu\nu}~\partial_{\nu}\bar{\lambda}^{\prime}_{l})
\nonumber \\
+ f^{(2a)}_{jkl}~\partial^{\nu}v^{\mu}_{k}~\partial_{\mu}v_{l\nu}
+ f^{(2c)}_{jkl}~\epsilon_{\mu\nu\rho\sigma}~
\partial^{\mu}v^{\nu}_{k}~\partial^{\rho}v_{l}^{\sigma}
\nonumber \\
+ f^{(4a)}_{jkl}~
(\lambda^{\prime}_{k}~\sigma^{\mu}~\partial_{\mu}\bar{\lambda}_{l}^{\prime})
+f^{(4b)}_{jkl}~
(\partial_{\mu}\lambda^{\prime}_{k}~\sigma^{\mu}~\bar{\lambda}_{l}^{\prime})
\nonumber \\
- 2~g^{(2a)}_{jkl}~v^{\mu}_{k}~\partial_{\mu}d_{l}
- 2~g^{(2b)}_{jkl}~\partial_{\mu}v^{\mu}_{k}~d_{l} 
- 2~g^{(2c)}_{jkl}~d_{k}~d_{l}
\eea

\bea
A^{\prime}_{j} = f^{(6b)}_{ljk}~
(\partial_{\mu}\chi_{k}~\sigma^{\mu\nu}~\partial_{\nu}\chi_{l})
+ f^{(6d)}_{ljk}~
(\partial_{\mu}\bar{\chi}_{k}~\bar{\sigma}^{\mu\nu}~\partial_{\nu}\bar{\chi}_{l})
\nonumber \\
+ f^{(9e)}_{ljk}~(\partial_{\mu}\chi_{l}~\sigma^{\mu}~\bar{\lambda}^{\prime}_{k})
+ f^{(9f)}_{ljk}~(\chi_{l}~\sigma^{\mu}~\partial_{\mu}\bar{\lambda}^{\prime}_{k})
\nonumber \\
+ f^{(9g)}_{ljk}~(\partial_{\mu}\lambda^{\prime}_{l}~\sigma^{\mu}~\bar{\chi}^{\prime}_{k})
+ f^{(9h)}_{ljk}~(\lambda^{\prime}_{l}~\sigma^{\mu}~\partial_{\mu}\bar{\chi}^{\prime}_{k})
\nonumber \\
+ (- f^{(10b)}_{kjl} + f^{(10d)}_{jkl})~\partial_{\mu}\phi_{k}~v_{l}^{\mu}
- f^{(10b)}_{kjl}~\phi_{k}~\partial_{\mu}v^{\mu}_{l}
\nonumber \\
+ (f^{(10c)}_{jkl} - f^{(10c)}_{kjl})~\partial_{\mu}\phi^{\dagger}_{k}~v_{l}^{\mu}
- f^{(10c)}_{kjl}~\phi^{\dagger}_{k}~\partial_{\mu}v^{\mu}_{l}
\nonumber \\
+ f^{(11b)}_{kjl}~\phi_{k}~d_{l}
+ 2~f^{(11c)}_{jkl}~\phi^{\dagger}_{k}~d_{l}
\nonumber \\
+ f^{(12b)}_{lkj}~(\lambda^{\prime}_{k}~\lambda_{l}^{\prime})
+ f^{(12d)}_{lkj}~(\bar{\lambda}^{\prime}_{k}~\bar{\lambda}_{l}^{\prime})
\eea

\bea
B^{\prime}_{j} = - f^{(6a)}_{ljk}~
(\partial_{\mu}\chi_{l}~\sigma^{\mu\nu}~\partial_{\nu}\chi_{k})
- f^{(6c)}_{ljk}~
(\partial_{\mu}\bar{\chi}_{l}~\bar{\sigma}^{\mu\nu}~\partial_{\nu}\bar{\chi}_{k})
\nonumber \\
+ f^{(9a)}_{ljk}~(\partial_{\mu}\chi_{l}~\sigma^{\mu}~\bar{\lambda}^{\prime}_{k})
- f^{(9b)}_{ljk}~(\chi_{l}~\sigma^{\mu}~\partial_{\mu}\bar{\lambda}^{\prime}_{k})
\nonumber \\
- f^{(9c)}_{ljk}~(\partial_{\mu}\lambda^{\prime}_{l}~\sigma^{\mu}~\bar{\chi}^{\prime}_{k})
+ f^{(9d)}_{ljk}~(\lambda^{\prime}_{l}~\sigma^{\mu}~\partial_{\mu}\bar{\chi}^{\prime}_{k})
\nonumber \\
+ (- f^{(10a)}_{jkl} + f^{(10a)}_{kjl})~\partial_{\mu}\phi_{k}~v_{l}^{\mu}
+ f^{(10a)}_{kjl}~\phi_{k}~\partial_{\mu}v^{\mu}_{l}
\nonumber \\
+ (- f^{(10b)}_{jkl} + f^{(10d)}_{kjl})~\partial_{\mu}\phi^{\dagger}_{k}~v_{l}^{\mu}
+ f^{(10d)}_{kjl}~\phi^{\dagger}_{k}~\partial_{\mu}v^{\mu}_{l}
\nonumber \\
- 2 f^{(11a)}_{jkl}~\phi_{k}~d_{l}
- f^{(11b)}_{jkl}~\phi^{\dagger}_{k}~d_{l}
\nonumber \\
- f^{(12a)}_{lkj}~(\lambda^{\prime}_{k}~\lambda_{l}^{\prime})
- f^{(12c)}_{lkj}~(\bar{\lambda}^{\prime}_{k}~\bar{\lambda}_{l}^{\prime})
\eea

\bea
X_{j} = - 2 f^{(1a)}_{kjl}~
\partial_{\mu}C_{k}~\sigma^{\mu\nu}~\partial_{\nu}\lambda^{\prime}_{l}
\nonumber \\
- 4 f^{(6a)}_{jkl}~\sigma^{\mu\nu}~\partial_{\nu}\chi_{k}~\partial_{\mu}\phi_{l}
- 4 f^{(6b)}_{jkl}~\sigma^{\mu\nu}~\partial_{\nu}\chi_{k}~\partial_{\mu}\phi^{\dagger}_{l}
\nonumber \\
+ 2 (- f^{(7a)}_{jkl} - f^{(7b)}_{jkl} + f^{(7c)}_{jkl} + f^{(7d)}_{jkl})~ \sigma^{\mu}~\partial_{\mu}\partial_{\nu}\bar{\chi}_{k}~v^{\nu}_{l}
\nonumber \\
+ 2 (- f^{(7a)}_{jkl} + f^{(7c)}_{jkl})~ \sigma^{\mu}~\partial_{\nu}\bar{\chi}_{k}~\partial_{\mu}v^{\nu}_{l}
\nonumber \\
+ 2 (- f^{(7b)}_{jkl} + f^{(7c)}_{jkl})~ \sigma^{\mu}~\partial_{\mu}\bar{\chi}_{k}~\partial_{\nu}v^{\nu}_{l}
+ 2 f^{(7c)}_{jkl}~\sigma^{\mu}~\bar{\chi}_{k}~\partial_{\mu}\partial_{\nu}v^{\nu}_{l}
\nonumber \\
- 2 f^{(7e)}_{jkl} \epsilon_{\mu\nu\rho\lambda} 
\sigma^{\nu} \partial^{\rho}\bar{\chi}_{k} \partial^{\mu}v_{l}^{\lambda}
\nonumber \\
+ 2 (- f^{(8a)}_{jkl} + f^{(8b)}_{jkl})~ \sigma^{\mu}~\partial_{\mu}\bar{\chi}_{k}~d_{l}
- 2 f^{(8a)}_{jkl}~ \sigma^{\mu}~\partial_{\mu}\bar{\chi}_{k}~d_{l}
\nonumber \\
+ 2 (f^{(9a)}_{jkl} + f^{(9b)}_{jkl})~
\sigma^{\mu}~\partial_{\mu}\bar{\lambda}^{\prime}_{k}~\phi_{l}
+ 2 f^{(9a)}_{jkl}~\sigma^{\mu}\bar{\lambda}^{\prime}_{k}~\phi_{l}
\nonumber \\
+ 2 (- f^{(9e)}_{jkl} + f^{(9f)}_{jkl})~
\sigma^{\mu}~\partial_{\mu}\bar{\lambda}^{\prime}_{k}~\phi^{\dagger}_{l}
- 2 f^{(9e)}_{jkl}~\sigma^{\mu}\bar{\lambda}^{\prime}_{k}~\phi^{\dagger}_{l}
\nonumber \\
+ (- 2 f^{(15a)}_{jkl} + 2 f^{(15b)}_{jkl})~
\sigma_{\mu\nu}~\partial^{\mu}\lambda^{\prime}_{k}~v^{\nu}_{l}
- 2 f^{(15a)}_{jkl}~\sigma_{\mu\nu}\lambda^{\prime}_{k}~\partial^{\mu}v^{\nu}_{l}
\nonumber \\
+ 2~(- f^{(15e)}_{jkl} + f^{(15f)}_{jkl})~
\partial_{\mu}\lambda^{\prime}_{k}~v^{\mu}_{l}
- 2 f^{(15e)}_{jkl}~\lambda^{\prime}_{k}~\partial_{\mu}v^{\mu}_{l}
\nonumber \\
+ 2 f^{(16a)}_{jkl} \lambda^{\prime}_{k}~d_{l}
\eea

\bea
\bar{X}^{\prime}_{j} = 2 f^{(1b)}_{kjl}~
\partial_{\mu}C_{k}~\bar{\sigma}^{\mu\nu}~\partial_{\nu}\bar{\lambda}^{\prime}_{l}
\nonumber \\
+ 4 f^{(6c)}_{jkl}~\partial_{\nu}\bar{\chi}_{k}\bar{\sigma}^{\mu\nu}~\partial_{\mu}\phi_{l}
+ 4 f^{(6d)}_{jkl}~
\partial_{\nu}\bar{\chi}_{k}\bar{\sigma}^{\mu\nu}~\partial_{\mu}\phi^{\dagger}_{l}
\nonumber \\
+ 2 (- f^{(7a)}_{kjl} - f^{(7b)}_{kjl} + f^{(7c)}_{kjl} + f^{(7d)}_{kjl})~ \partial_{\mu}\partial_{\nu}\bar{\chi}_{k}\sigma^{\mu}~v^{\nu}_{l}
\nonumber \\
+ 2 (- f^{(7a)}_{kjl} + f^{(7d)}_{kjl})~
\partial_{\mu}\chi_{k} \sigma^{\mu}~\partial_{\nu}v^{\nu}_{l}
\nonumber \\
+ 2 (- f^{(7b)}_{kjl} + f^{(7d)}_{kjl})~
\partial_{\mu}\chi_{k}~\sigma^{\nu}~\partial_{\nu}v^{\mu}_{l}
+ 2 f^{(7d)}_{kjl}~\chi_{k}~\sigma^{\mu}~~\partial_{\mu}\partial_{\nu}v^{\nu}_{l}
\nonumber \\
- 2 f^{(7e)}_{kjl} \epsilon_{\mu\nu\rho\lambda} 
\partial^{\mu}\chi_{j} \sigma^{\nu} \partial^{\rho} v_{l}^{\lambda}
\nonumber \\
+ 2 ( f^{(8a)}_{kjl} - f^{(8b)}_{kjl})~\sigma^{\mu}~\partial_{\mu}\bar{\chi}_{k}~d_{l}
- 2 f^{(8b)}_{jkl}~\partial_{\mu}\chi_{k}~\sigma^{\mu}~d_{l}
\nonumber \\
+ 2 (f^{(9c)}_{kjl} - f^{(9d)}_{kjl})~
\partial_{\mu}\lambda^{\prime}_{k}~\sigma^{\mu} \phi_{l}
- 2 f^{(9d)}_{kjl}~\lambda^{\prime}_{k}~\sigma^{\mu}~\partial_{\mu}\phi_{l}
\nonumber \\
+ 2 (f^{(9g)}_{kjl} - f^{(9h)}_{kjl})~
\partial_{\mu}\lambda^{\prime}_{k}~\sigma^{\mu} \phi^{\dagger}_{l}
- 2 f^{(9h)}_{kjl}~\lambda^{\prime}_{k}~\sigma^{\mu}~\partial_{\mu}\phi^{\dagger}_{l}
\nonumber \\
+ 2 (f^{(15c)}_{jkl} - f^{(15d)}_{kjl})~
\partial^{\mu}\bar{\lambda}^{\prime}_{k}~\bar{\sigma}_{\mu\nu} v^{\nu}_{l}
+ 2 f^{(15c)}_{jkl}~
\bar{\lambda}^{\prime}_{k}~\bar{\sigma}_{\mu\nu}~\partial^{\mu}v^{\nu}_{l}
\nonumber \\
- 2 (f^{(15g)}_{jkl} - f^{(15h)}_{kjl})~
\partial_{\mu}\bar{\lambda}^{\prime}_{k}~v^{\mu}_{l}
+ 2 f^{(15g)}_{jkl}~\bar{\lambda}^{\prime}_{k}~\partial_{\mu}v^{\mu}_{l}
\nonumber \\
- 2 f^{(16b)}_{jkl} \bar{\lambda}^{\prime}_{k}~d_{l}
\eea

If we make a general ansatz
\be
T^{\mu} = u_{j} T^{\mu}_{j} + (\partial_{\nu}u_{j}) T^{\mu\nu}_{j}
+ (\partial_{\nu}\partial_{\rho}u_{j}) T^{\mu\nu\rho}_{j}
\ee
we can prove, as in the Yang-Mills case, that we must have in fact 
\be
A_{j} = 0  \quad A_{j}^{\prime} = 0 \quad B_{j} = 0  \quad B_{j}^{\prime} = 0 \quad
X_{j} = 0  \quad X_{j}^{\prime} = 0. 
\ee
In particular we have from the coefficient
of
$
\partial^{\nu}v^{\mu}_{k}~\partial_{\mu}v_{l\nu}
$
that
\be
f^{(13)}_{jkl} = - f^{(13)}_{kjl};
\label{a7}
\ee
together with 
\be
g^{(1b)}_{jkl} = g^{(1b)}_{kjl} ;
\label{s1}
\ee
this amounts to the total antisymmetry of the expression
$
f^{(13)}_{jkl}
$
and the Yang-Mills solution emerges. We also have
\be
g^{(12a)}_{jkl} = 2 i~f^{(14)}_{jlk} \qquad
g^{(12b)}_{jkl} = 2 i~f^{(14)}_{ljk}
\ee
and the second solution emerges.

One can make a double check of the computation as follows: one eliminates only the terms of the type
$
\tilde{F}, \tilde{G}
$
and do not use the co-boundaries.  So, we work with the whole set of 132 terms of type
$
F, G.
$
As a result we find 35 solutions of which 33 are trivial and the other two are those already obtained above.

\end{document}